\documentclass[10pt,journal,compsoc]{IEEEtran}

\ifCLASSOPTIONcompsoc
  \usepackage[nocompress]{cite}
\else
  \usepackage{cite}
\fi

\ifCLASSINFOpdf
  \usepackage[pdftex]{graphicx}
\else
  \usepackage[dvips]{graphicx}
\fi

\ifCLASSOPTIONcompsoc
\usepackage[caption=false,font=footnotesize,labelfont=sf,textfont=sf]{subfig}
\else
  \usepackage[caption=false,font=footnotesize]{subfig}
\fi

\usepackage{amsmath,amssymb,amsfonts}
\usepackage{algorithmic}
\usepackage{textcomp}
\usepackage{bmpsize}
\usepackage{xcolor}
\usepackage{lipsum}
\usepackage{array}
\usepackage[colorlinks=true,urlcolor=black]{hyperref}
\usepackage[utf8]{inputenc}
\usepackage{dblfloatfix} 
\usepackage[linesnumbered, ruled]{algorithm2e}
\usepackage{cuted}
\usepackage{amsthm}
\usepackage{url}

\renewcommand{\IEEEbibitemsep}{0pt plus 0.5pt}
\makeatletter
\IEEEtriggercmd{\reset@font\normalfont\fontsize{7.9pt}{8.40pt}\selectfont}
\makeatother
\IEEEtriggeratref{1}

\usepackage[compact]{titlesec}
\titlespacing{\section}{0pt}{2ex}{1ex}
\titlespacing{\subsection}{0pt}{1ex}{0ex}
\titlespacing{\subsubsection}{0pt}{0.5ex}{0ex}

\theoremstyle{definition}

\begin{document}
 
  \title{Privacy Preserving Release of Mobile Sensor Data}
 
 \author{
	\IEEEauthorblockN{
	Rahat Masood, \IEEEauthorrefmark{1}
        Wing Yan Cheng, \IEEEauthorrefmark{1}
        Dinusha Vatsalan, \IEEEauthorrefmark{2}
        Deepak Mishra, \IEEEauthorrefmark{1}
        Hassan Jameel Asghar, \IEEEauthorrefmark{3}
        Mohammed Ali Kaafar \IEEEauthorrefmark{3}
	}\\ 
	
		\IEEEauthorblockA{
		\IEEEauthorrefmark{1} UNSW, Sydney 
		\IEEEauthorrefmark{2} Data61, CSIRO, 
		\IEEEauthorrefmark{2} Macquarie University,  Australia \\  
		Email:  \{firstname.lastname\}@unsw.edu.au,
		\{firstname.lastname\}@dat61.csiro.au,
		\{firstname.lastname\}@mq.edu.au \\
	}
	
}

\IEEEtitleabstractindextext{%
\begin{abstract}

Sensors embedded in mobile smart devices can monitor users' activity with high accuracy to provide a variety of services to end-users ranging from precise geolocation, health monitoring, and handwritten word recognition. However, this involves the risk of accessing and potentially disclosing sensitive information of individuals to the apps that may lead to privacy breaches. In this paper, we aim to minimize privacy leakages that may lead to user identification on mobile devices through user tracking and distinguishability while preserving the functionality of apps and services. We propose a privacy-preserving mechanism that effectively handles the sensor data fluctuations (e.g., inconsistent sensor readings while walking, sitting, and running at different times) by formulating the data as time-series modeling and forecasting. The proposed mechanism also uses the notion of correlated noise-series against noise filtering attacks from an adversary, which aims to filter out the noise from the perturbed data to re-identify the original data. Unlike existing solutions, our mechanism keeps running in isolation without the interaction of a user or a service provider. We perform rigorous experiments on three benchmark datasets and show that our proposed mechanism limits user tracking and distinguishability threats to a significant extent compared to the original data while maintaining a reasonable level of utility of functionalities. In general, we show that our obfuscation mechanism reduces the user trackability threat by 60\% across all the datasets while maintaining the utility loss below 0.5 Mean Absolute Error (MAE). We also observe that our mechanism is more effective in large datasets. For example, with the GPS dataset containing 464,033 samples, the user trackability risk is reduced by 70\%, and with the Swipes dataset, the distinguishability risk is reduced by 60\% on average while the utility loss is below 0.5 MAE.

\end{abstract}
}

\maketitle
\IEEEdisplaynontitleabstractindextext
\IEEEpeerreviewmaketitle

\IEEEraisesectionheading{\section{Introduction}\label{intro}}

\vspace{-2.5mm}
With the increasing use of mobile devices, a vast amount of temporal sensory data is generated by individuals through the use of mobile apps and collected by service providers through sensors such as microphones, cameras, motion accelerators, gyroscopes, touch gestures, and GPS trajectories. In many cases, service providers or the app owners capture and analyze raw sensor data in order to perform the required app functionalities or for other analytics purposes such as improving user experience~\cite{zhan2012accurate, nyt-teen}. 

For instance, a messenger app needs raw sensors data, such as \textit{x} and \textit{y} coordinates, to recognize letters entered by a user on a touch screen. This app can also capture the user's writing behavior e.g., how a user slides his fingers while typing messages. Similarly, a health monitoring app needs motion sensor data to detect activities (e.g., sitting, walking, and running) of a person. At the same time, it can identify the user's unique walking style for purposes such as improved user experiences. Moreover, there are some apps that are solely developed for analytics purposes; for example, \textit{Appsee} \cite{appsee}, performs an in-depth analysis of users' behavior using scroll and touch gestures collected through sensors data.

The retrieval of such fine-grained information is not always necessary for the app's functionality as it may lead to privacy breaches such as user tracking or identification~\cite{das2016tracking, Dey2014, Miluzzo2012, masood2018touch, berend2018there, das2018web, cnbc-news} Abstractly, we consider the scenario where a user interacts with an app on his/her mobile device either explicitly, i.e., via some gestures (swipe, tap, write), or implicitly, e.g., by giving GPS access to an app. Data related to these interactions (e.g. raw x, y coordinates, finger pressure, etc.) is sent to the app's remote server, to acquire service in return. While some of the data is necessary to provide the desired functionality, not all information is required.   Consider, for instance, the following examples:
\begin{itemize}

\item Alice would like to use an app that transforms her handwritten letters into a notebook document. The app needs to be able to recognize letters, however it does not need to extract her unique writing style as it may profile Alice and link her across different sessions~\cite{masood2018touch}.

\item George would like to use a document reader to read articles on his mobile. The app needs to know where the user has tapped or scrolled but does not need to know any behavioral patterns about George, e.g., speed of swipes, duration, or the pressure exerted by his finger, as it may lead to tracking \cite{try}. 
\end{itemize}

Researchers have proposed solutions to overcome user identification and tracking issues from mobile sensory data~\cite{das2016tracking, das2018every}. However, these solutions are not generic as they do not hold for all types of mobile sensors data types such as touch sensors. Few recent studies~\cite{raval2019olympus, malekzadeh2018mobile} either require developers to hookup the private mechanism with their apps or depend on a user to train the learning mechanism. In addition, previous solutions did not discuss the fluctuations\footnote{The fluctuation occurs when mobile devices are carried and used in different situations such as sitting, walking, or running.} of mobile sensor data that may negatively impact the privacy and utility results of an obfuscation mechanism. Finally, all existing works are evaluated on a single privacy metric (i.e., user identification through distinguishability) and did not discuss the privacy notion of user trackability across different sessions.

We define \textit{user distinguishability} as an ability of an adversary to uniquely distinguish and identify a user from the data of all users. We refer \textit{user trackability} as an ability of an adversary to link a current session (a specific period) to a user based on his/her data from previous sessions. A user that is distinguishable among all the other users, is not necessarily trackable across different sessions, as his behavior to perform activities may change with time. On the contrary, a trackable user might be distinguishable among users because of the uniformity (same pattern) in the observed data during tracking. For example, mobile sensor data captured at a particular time shows that the user is distinguishable from other users. However, the same user reflects continuously changing behavior across various time intervals and is thus, not trackable. Through this work, \textit{we investigate whether we can overcome the privacy issues of user tracking or distinguishability, associated with mobile sensory data release, by providing a generic on-device privacy preserving framework, addressing all the abovementioned limitations.}  We make the following contributions in this paper:

\begin{enumerate}
    \item We design an on-device privacy-preserving mechanism that minimizes the leakages of private information of a user before releasing it to a service provider, whilst maintaining the intended utility of an app/service. Our proposed mechanism overcomes the drawbacks of previous solutions such as application specifity, user interaction with a privacy preserving mechanism, and trust issues with service providers. We evaluate our mechanism against two privacy metrics, \textit{trackability} and \textit{distinguishability}, that quantify risks involving user tracking and identification, from mobile sensory data.

    \item We formulate our problem as time-series data modelling that utilizes time-series methods such as TBATS\footnote{TBATS: Trigonometric seasonality, Box-Cox transformation, ARIMA errors, Trend, Seasonal component} modeling and forecasting~\cite{de2011forecasting}, Dickey and Fuller (ad-fuller) test~\cite{dickey1981likelihood}, and box-cox transformation~\cite{sakia1992box}. 
      By using time-series data modeling, our mechanism offers two levels of data privacy protection; first by replacing original time-series data with forecasted data and second by adding correlated noise to the forecasted data. The use of time-series forecasting methods also provides a good balance between privacy and utility when the sensory data is fluctuating. Particularly, the ad-fuller test checks whether the upcoming time-series is stable (fluctuating)~\cite{dickey1981likelihood}. If the series is unstable, then box-cox transformation is used to make the series stable~\cite{sakia1992box}. The stabilized time-series is then fed into TBATS modeling to make accurate forecasts on the sensory data.  Finally, the proposed mechanism adds correlated noise to the forecasted time-series in order to provide privacy guarantees.

    \item Our framework is resilient against noise filtering attacks by an adversary since it generates correlated noise-series that is added to the forecasted time-series data. Correlated noise implies that the correlation of the noise series between the original series is high and therefore, it prevents noise filtering attack from adversaries~\cite{wang2017cts}.

    \item We empirically show the effectiveness of our mechanism in three different scenarios: (i) Handwriting by Stylus and (ii) Touch Swipes. We perform a comprehensive experimental study by evaluating our framework in terms of trackability, indistinguishability, utility, and efficiency. We show that trackability and distinguishability threats are reduced significantly while maintaining a reasonable level of utility. In general, our obfuscation mechanism shows a reduction in the threats of tracking and distinguishing users on average by 60\% and 35\%, respectively. Similarly, the handwritten letters recognition dataset shows a 65\% reduction in distinguishability with only 0.53 (MAE) of utility loss. Moreover, our experimental results show that a good balance between privacy and utility is possible by fine-tuning the noise addition parameter values. These parameters can be changed anytime, through OS updates, to satisfy the requirements of privacy-utility tradeoff.
\end{enumerate}

The rest of the paper is organized as follows: Section \ref{prelim} describes the scenario of mobile sensory data flow and defines the threat model of users' information leakage from the mobile devices. Section \ref{method_mob} presents our privacy-preserving mechanism that includes time-series data training and time-series privacy preservation as main phases. Section \ref{sec:eval_mob} and \ref{res_sec} discusses experimental setup and results of our study, respectively. Section~\ref{dis} summarizes our findings and Section \ref{related_work} describes related work. Finally, Section \ref{conclusion_mob_pri} discusses the limitations and future directions, along with the conclusion.

\section{Preliminaries}
\label{prelim}

In this section, we begin with the background describing the mobile data flow then, we define the threat model of our study.

\subsection{Background}
\label{background}
Consider a scenario in which a user has installed an app on his/her mobile device. The app is interactive i.e. the user performs some actions in return for a functionality offered by the app. User actions are interpreted via readings from the mobile device sensors accessible via the API, e.g., touch gestures. As explained before, the app does not need information from all sensors or even all information from a sensor for its \emph{correct} functionality~\cite{masood2018touch}\footnote{Most of the apps have their detailed logic implemented at the remote server-side, which means information needs to flow from the user's device to the remote server.}. For instance, a song identification app only needs to know that the user has ``tapped'' on the button issuing the command to recognize the song played in the background.  However, it does not need to know fine-grained details such as the pressure the user exerts to perform the tap or the area covered.

Ideally, information accessible to the app could be filtered by the OS, such that only information necessary for the app's correct functionality is communicated to its server. However, this is arguably impractical due to various reasons. First, if we know what information is required for the correct functionality of the app, it is difficult to deduce the required granularity of information. Note that our concern is tied to algorithmic or programmatic detection of these information; a human expert might be able to deduce this, but this is far from a practical solution not just due to the sheer volume of third-party apps. One can likewise not rely on app developer(s) to be honest about these details due to obvious commercial benefits. Secondly, the app might be programmed to expect fine-grained information from the user, in which case withholding some of this information will make the app work improperly (even though not all the information is required for the correct functionality of the app).

Nevertheless, we also note that the necessary information may itself be privacy-intrusive. For instance, in the location example above, just the time at which the user uses the recommendation app and the approximate location of the user may be enough to identify the user (and hence track the user). Therefore, our goal is to construct a privacy-preserving mechanism that mimics the ideal setting outlined above by focusing on the information available through the sensor readings. To make our treatment more precise, we define a few terms. \vspace{-2mm}

\begin{enumerate}
    \item \emph{Users:} We assume an app has a finite set $\mathcal{U}$ of $k$ users. A user in $\mathcal{U}$ is denoted as $u$.
    \item \emph{Session:} A session is defined as the collection of activities performed by a user between he/she opening and closing the app.
    \item \emph{Sensor readings:} For each session, there is a finite sequence of readings from each of the sensors available through the device via an API. We assume a fixed sampling rate at which the data can be read via the API. We can think of multiple functions being available via the API to access different information about the sensors. For instance, in the case of a touch API, one function conveys whether a tap event has been performed, whereas another function can be called to know the pressure information. Without complicating things, we will assume a finite set $\mathcal{S}$ of possible types of sensor information, which can be assumed to correspond to different functions available through the API.
    \item \emph{Gesture:} A gesture $g$ is a contiguous subsequence of sensor readings of non-zero length. We define the set of all gestures in a session by $\mathcal{G}$.\footnote{A gesture may be explicit, e.g., a user swipe, or implicit, e.g., GPS coordinates being communicated to the app.}
    \item \emph{Functionality:} The functionality of the app is defined as a function $f : \mathcal{G} \rightarrow \mathcal{E}$ that takes a gesture $g \in \mathcal{G}$ as input and outputs a subset $E \in \mathcal{E}$ of events, where $\mathcal{E}$ is an arbitrary finite set of events.

\end{enumerate}

\subsection{Threat Model}
\label{sub:threat}

Our threat model considers a service provider as an untrustworthy entity that collects sensors' readings with the purpose to uniquely identify or track users across sessions. We quantify such threat using two privacy risk metrics i.e., \textit{trackability} and \textit{distinguishability}.

\subsubsection{Trackability} 
To quantify the privacy risk of user tracking through the trackability metric, consider a user $u$, who performs a certain set of gestures $\mathcal{G}$ in a session. The gestures produce a set of sensor readings $\mathcal{S}$ such that some of the readings are necessary to achieve a functionality whereas, some of the readings are unnecessarily collected (not relevant to functionality) to track users in the future sessions. We refer latter as `public information ($Pub$)' and former as `private information ($Pri$)'. For-example, $Pub$ corresponds to the information that is necessary in the recognizing the letter or word written by a user, e.g. \{`hello', `hi', `howdy'\}, while $Pri$ is user's writing style or behaviour of writing. Both $Pub$ and $Pri$ are extracted from sensor readings $\mathcal{S}$ such that $(Pub, Pri) \in \mathcal{S}$. By collecting $\mathcal{S}$ from multiple sessions, an adversary can deduce unique/identifiable information about a user that could lead to tracking or linking him to new sessions. 

We call this threat as \textit{trackability}, which is the ability of an adversary to link two or more sessions from the same user with high likelihood between the sessions' data.

\subsubsection{Distinguishability}
This refers to the ability of an adversary to distinguish data of one user from all other users. To understand distinguishability, consider a scenario where an adversary has a collection of gestures from different users e.g., handwritten data. The goal of an adversary is to distinguish one user's data from others by quantifying the amount of unique information available from the dataset. In the best-case scenario, all users look similar in the data, and thus, the adversary is unable to uniquely identify any user. In the worst case, each user in the dataset is different and unique, where the adversary can easily distinguish one user from all others. In technical terms, if the probability of linking a gesture $g$ to a user $u$ is higher than linking to all users in $\mathcal{U} \setminus \{u\}$, then the user $u$ is highly distinguishable.

The difference between trackability and distinguishability is that the former considers the uniformity (in behaviour) of the user across different sessions, while the latter considers the uniqueness of a user among other users. Uniformity and uniqueness of users' data are the key information that adversaries can use to identify and learn about an individual~\cite{masood2018incognito}.

\section{Methodology}
\label{method_mob}

In this section, we first provide an overview of the proposed idea and a sketch of our system model, and then describe our privacy-preserving mechanism that includes time-series data training and time-series privacy preservation as main phases.

\subsection{Proposed Idea}

Our main idea for the proposed method is as follows. 
\begin{itemize}
    \item 
    Each activity performed by a user on a mobile device is categorized into a specific gesture (e.g. for hand-writing dataset we would like to determine the which character is represented by a given time-series) using a data clustering process (cf. Section~\ref{frame-training}).

    \item 
    We represent each sample of the gesture as a time-series. We append the time-series of multiple instances of the same gesture type one after the other, that results in a ``periodic'' time-series, where each period corresponds to the time of each gesture instance.

    \item 
    We use time-series modeling and forecasting to predict a typical instance of the user's gesture. This typical gesture will be used whenever the user intends to input the same gesture type. 
    
    \item 
    While the forecasted time-series may mitigate some privacy risks by removing fine-grained details contained in gestures, it is highly likely that the forecasted gestures show some common characteristics of a user, thus making it easy to distinguish from other users. Therefore, we add noise to the predicted gesture to make it harder for an adversary to distinguish two predicted gestures, whether the gestures are from the same or different users.
    
    \item 
    The noise added to the gesture is correlated since successive points in the gesture are expected to be correlated. This ensures that an adversary cannot remove the noise by using the correlation information. 
    
\end{itemize}

To illustrate our solution through time-series, consider the following example:

\textbf{Example 1:} A user $u$ plays a game (e.g. 2048\footnote{https://github.com/gabrielecirulli/2048}) on a mobile and performs gestures such as \textit{left, right, up,} and \textit{down} swipes. Each gesture $g$ invokes a function from touch sensor API that collects raw sensor readings e.g. \textit{x-position, y-position,} and \textit{finger pressure} such that $\mathcal{S}[\text{pressure}, \text{x}, \text{y}] \in g$. Now we formulate each raw feature in a sensor reading as a single (uni-variate) time-series such that $\mathcal{S}[\text{pressure}, \text{x-pos}, \text{y-pos}] = [X_{p}, X_{x-pos}, X_{y-pos}]$. Here, $[X_{p}, X_{x-pos}, X_{y-pos}]$ are three time-series of a sensor-reading $\mathcal{S}$.

\textbf{Remark 1:} Throughout this paper, we refer time-series to a univariate time-series of a single gesture $g$ at a time interval of $t$. We refer time-series data to multiple instances of user gestures $\mathcal{G}$.

\subsection{System Overview}

\begin{figure}[!t]%
    \centering
    \includegraphics[scale= 0.4, keepaspectratio]{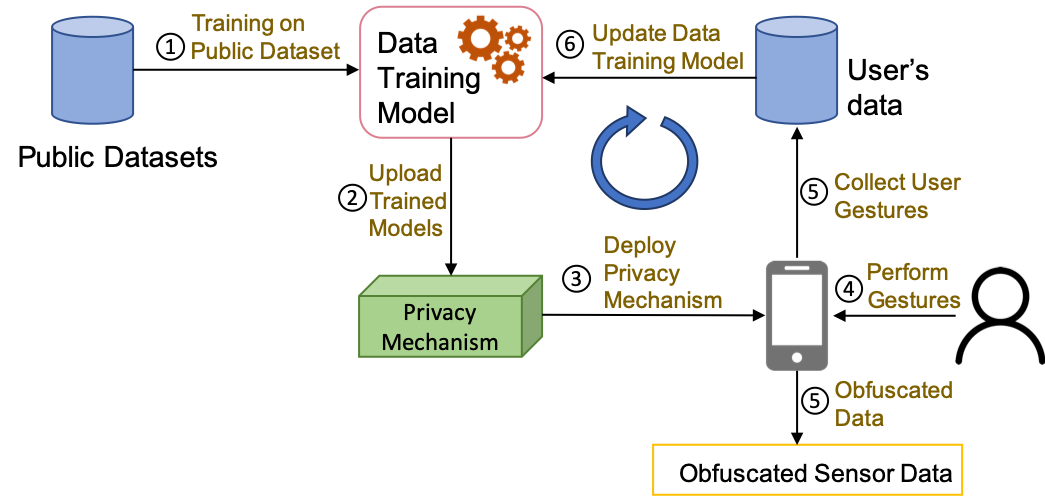}
    \caption{System Overview of Proposed Mechanism}
    \label{fig:sys_overview}
\end{figure}

Our proposed privacy-preserving mechanism is deployed in a device operating system (OS) such that the time-series data from any app must be first obfuscated at an OS level and then sent to a remote (mobile app) server\footnote{In a typical setting, a device OS receives numerous out-going records every day which are sent to designated servers at a fixed sampling rate. Depending on restrictions and rules deployed at the OS level, the data can be sent with or without pre-processing.}. The overall working mechanism is as follows and also illustrated in Figure~\ref{fig:sys_overview}: First, our obfuscation mechanism is trained on a public dataset that contains sensor readings from various gestures so that when a user starts using a device, the privacy preservation level is maintained to some extent. This training data is labeled with a gesture type. Once trained, we deploy our obfuscation mechanism on a mobile OS. A user starts using a mobile phone while his/her gesture data are obfuscated at the OS level using the proposed mechanism. Once sufficient data is collected from a user, our mechanism then updates the trained models in specific intervals, for example, hourly or daily. After updating, our obfuscation method then starts preserving sensor data before sending to the app server. Next, we explain the time-series data training and privacy-preservation phases in detail.

\subsection{Privacy-Preserving Mechanism}
\begin{figure}[!t]
    \centering
\subfloat[\tiny In the time-series data training phase, 1) training data is input to a cluster method to generate trained cluster model, 2) the stability of each cluster is checked using ADF test, 3) unstable clusters are stabilized using Box-Cox transformation, 4) time-series data modeling is performed on stable clusters using  TABTS model, and 5) trained clusters and forecast models are stored in a mobile OS]{\label{fig:training}
      \includegraphics[scale= 0.3, keepaspectratio]{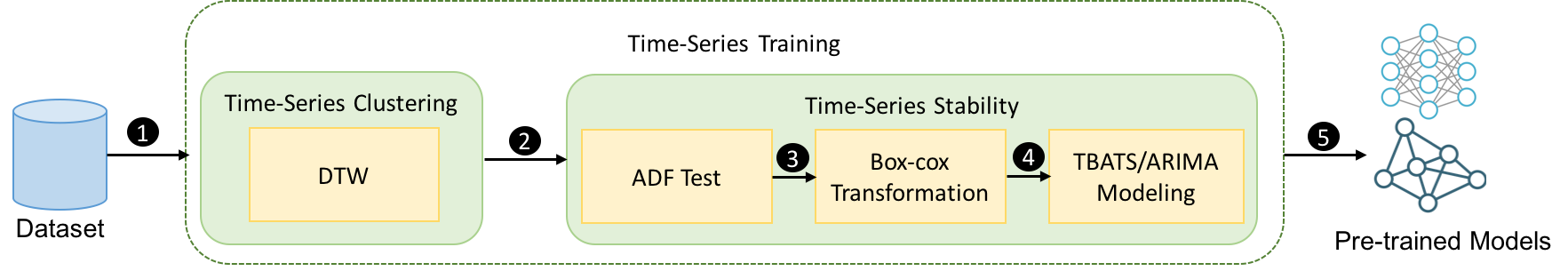}} \\
\subfloat[\tiny In the time-series privacy preservation phase, 1) trained models are deployed in a mobile OS and a new time-series (i.e. gesture performed at run-time) is input to the trained cluster model, 2) based on the selected cluster, time-series forecasting is performed, 3) forecasted time-series is then transformed back to original data scale using inverse Box-Cox transformation, 4) auto-correlation matrix is calculated for transformed time-series, 5) auto-correlation coefficients and white Gaussian noise are fed to LP filter to generate correlated noise, 6) correlated noise is added to the time-series and 7) the obfuscated time-series is released to the server] {\label{fig:privacy}
      \includegraphics[scale=0.27, keepaspectratio]{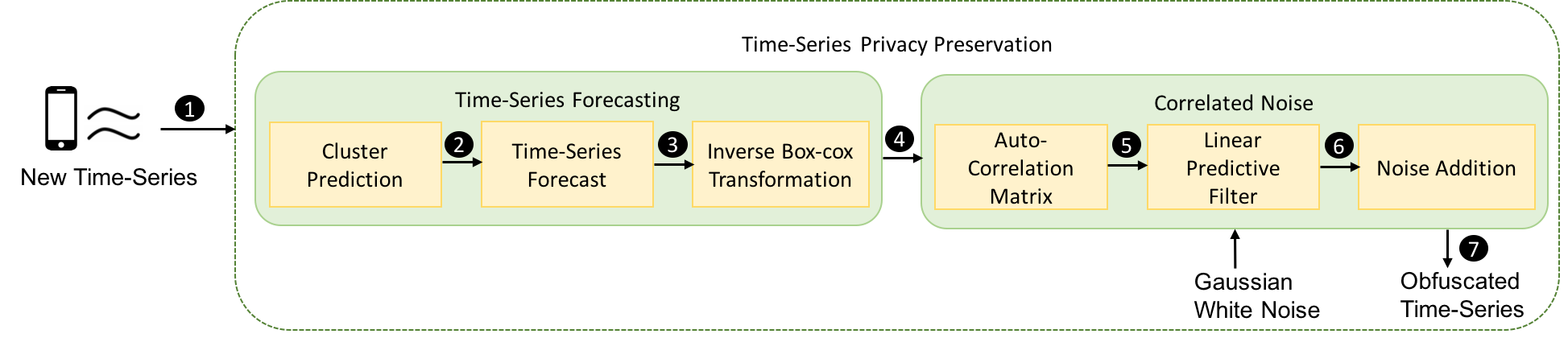}}
\caption{System Overview}
\label{fig:2}
\end{figure}

Our privacy-preserving mechanism consists of the following two phases: i) Time-Series Data Training (Figure \ref{fig:training}) and ii) Time-Series Privacy Preservation (Figure \ref{fig:privacy}). In this subsection, we discuss in detail these two phases and their components.

\subsubsection{Time-Series Data Training} 
\label{frame-training}
The Time-Series Data Training phase trains and deploys our obfuscation mechanism in the OS of a mobile device and also updates mechanism at regular intervals. In particular, the training phase has two components, namely: Time Series Data Clustering and Time Series Data Stability.

\textbf{Time-Series Data Clustering} refers to grouping similar time-series data into the same cluster and storing the trained cluster model in a device. In a typical setting, OS does not know the gesture type or the functionality of the sensor’s reading; it just sends data to the designated server. However, in our proposed obfuscation mechanism, it is necessary to distinguish gestures so as to produce accurate forecasted time series at run-time. The OS does not know exactly what the data is, for example, if the data is a swipe or a hand-written digit. It is, therefore, the requirement of the obfuscation mechanism to perform clustering to distinguish different gestures performed by the user. To do so, a public dataset that contains a set of different gestures, labelled with gesture type, is fed into a Dynamic Time Warping (DTW) model that groups similar time-series data (gestures).

\textbf{Time-Series Data Stability} refers to analyzing the fluctuations in a time-series and then making it stable for future predictions. The traditional classification and regression predictive modeling assumes that the summary statistics of observations are consistent \cite {trad-ML}. These assumptions can be easily violated in mobile sensor data due to fluctuations that occurred when mobile devices are used in different situations such as walking, driving, sitting, and standing\footnote{In the time-series terminology, this is known as the time series unstablity.}. Therefore, for mobile sensor data, it is necessary to generate a stable time-series in order to maintain high utility. For this purpose, we use ADF, a statistical test, to check the stability of time-series data.

Statistical tests can provide a quick check and confirmatory evidence whether the time-series data is stationary or non-stationary.

Once time-series data is determined to be non-stationary or unstable, a transformation such as Box-Cox transformation needs to be applied to make time-series data stable. Box-cox is a data transformation technique used to stabilize variance and make the data look like a normal distribution \cite{sakia1992box}. Since all data points in our dataset have positive values, we choose one-parameter version of Box-Cox transformation, which is defined as: \vspace{-2mm}
\begin{equation}
    y_i^{(\lambda)} =  \begin{cases} 
   \frac{y_i^{\lambda} - 1}{\lambda}, &  \text{if } \lambda \neq 0 \\
   \log (y_i),      &  \text{if }  \lambda = 0 ,
  \end{cases}
  \label{eq:box}
\end{equation}
where $y_i$ is a time-series observation at interval $i$ and $\lambda$ is a parameter with values ranging from -5 to 5.  All values of $\lambda$ are considered and the optimal value for the data is selected. The ``optimal value'' is the one which results in the best approximation of a normal distribution curve of a time-series.

Once stabilized, the time-series data is input into a forecasting model, TBATS~\cite{de2011forecasting}, to train the model. The TBATS model basically belongs to a general class of different models, such as random walk, random trend, seasonal and non-seasonal exponential smoothing and auto-regressive model, and it uses a systematic procedure for identifying the best model for any given time-series. The most important step in such type of modeling is to select values for parameters of various components such as seasonality, trends, moving average, etc. There exist systematic procedures that select suitable parameter values based on the time-series data \cite{hyndman2007automatic}. For our work, we use an auto-forecasting model that selects the appropriate parameter values~\cite{auto_forecast} and then fits the model on a time-series. Algorithm \ref{algo:1} illustrates the algorithmic description of the training phase. Once trained, our obfuscation mechanism starts forecasting time-series at run-time. 

\begin{algorithm}[!t]
\tiny
\DontPrintSemicolon
\SetKwInOut{Input}{Input}\SetKwInOut{Output}{Output}
\Input{$T$ = time-series data for training, \newline $\mathcal{U}$ = set of all users, \newline $F$ = set of time-series where each represents a feature in $T$}
\Output{$V$=pre-trained forefasting models , \newline $C$ = pre-trained clustering models}

\BlankLine

\For{$i\leftarrow 1$ \KwTo $\mathcal{U}$}{
\For{$f\leftarrow 1$ \KwTo $F$}{
Perform clustering on $T$ using Dynamic Time Wrapping (DTW) \\
Store resulted clustering model into $C$ and cluster labels $L$ \\

\For{$l\leftarrow 1$ \KwTo $L$}{
    Train TBATS model with stabilized time-series $T'$ to get a trained model $V$\\
    Store resulted TBATS model into $V$
    }
    }
}
\caption{Time-series Training (Offline Processing)}
\label{algo:1}
\end{algorithm}
\subsubsection{Time-Series Privacy Preservation}
\label{preservation_phase}
After training, the next phase is to preserve users' sensitive data at run-time by intercepting at OS and then send the perturbed data to a remote server. The two main components in this phase are: Time-Series Data Forecasting and Correlated Noise.

\textbf{Time-Series Data Forecasting}
is an additional layer between original and the perturbed data. The purpose of forecasting a time-series is to handle fluctuations in sensors data. Our obfuscation mechanism is trained on a certain set of data that has been previously stabilized, so it may not provide similar utility guarantees with real-time data or capable to handle fluctuation. It is thus, necessary to stabilize and forecast the data so to minimize the impact on utility.

In order to forecast, first the gesture type (e.g. a swipe, a tap, or a hand written letter) of an incoming series is determined using the pre-trained cluster model. As mentioned in Section~\ref{frame-training}, we use DTW model to cluster different gesture types in the trained dataset and then use the trained cluster model to predict the cluster of future data.
Once the cluster is selected, the trained forecasting model is called to predict the typical instance of a time-series (gesture). Next, we reverse the Box-Cox transformation that we have previously applied to the data, the purpose is to obtain forecasts on the original scale. The reverse Box-Cox transformation is given by: \vspace{-1.5mm}
\begin{equation}
    y_i =  \begin{cases} 
   (\lambda y_i^{(\lambda)} + 1)^ \frac{1}{\lambda}, &  \text{if } \lambda \neq 0 \\
   \exp (y_i^{(\lambda)}),      &  \text{if }  \lambda = 0
  \end{cases}
  \label{eq:inv-box}
\end{equation}
where $y_i^{(\lambda)}$ is the transformed data using parameter $\lambda$.
\textbf{Correlated Noise} is added to forecasted data to provide sufficient privacy guarantees. The purpose of adding correlated noise is to prevent the threat of re-identifying original series by filtering out noise from perturbed time-series data~\cite{wang2017cts}. Our correlated noise generation is based on the concept of linear prediction (LP) that is widely used in DSP. The purpose is to find linear filter coefficients based on the forecasted time-series such that when a random noise is passed through a filter, a noise correlated to forecasted data is generated\footnote{more detailed steps about creating linear filter is provided in the supplementary material}. 

\begin{equation}
\tiny
  \underbrace{ \begin{pmatrix}
  R_{x}[0] & R_{x}[1] & \cdots & R_{x}[N-1] \\
  R_{x}[1] & R_{x}[0] & \cdots & R_{x}[N-2] \\
  \vdots  & \vdots  & \ddots & \vdots  \\
  R_{x}[N-1] & R_{x}[N-2] & \cdots & R_{x}[0] 
 \end{pmatrix}}_{R_{x_{0:N}}}
 \underbrace{\begin{pmatrix}
  a_{0} \\
  a_{1} \\
  \vdots \\
  a_{N} 
 \end{pmatrix}}_{a}  =
\\
 \underbrace{\begin{pmatrix}
  R_{x}[1] \\
  R_{x}[2] \\
  \vdots \\
  R_{x}[N] 
 \end{pmatrix}}_{r} 
 \label{eq:corr_mat}
\end{equation}
which can be solved for the unknown coefficients, by setting \vspace{-1.5mm}
\begin{equation}
    a= R^{-1}_{x_{0:N}}r
    \label{coff}
\end{equation}

The matrix, $R_{x_{0:N}}$ is the auto-correlation matrix of the gesture instance, $x_{0:n}$, whose elements are $x[0]$ through $x[N-1]$. We may call this the $Nth$ order auto-correlation matrix for the gesture instance, $x[n]$. The coefficients $a_1, a_2,....a_N$ for $N$ order could be extracted by finding the values of $R_{{x}_{0:N}}$ and $r$ from Eq. \ref{coff}. It is to be noted that to find a linear filter coefficients with the best correlation effects, the auto-correlation matrix of any order from $1$ to $N$ should be tested. To do so, we use auto-correlation matrix from $1$ to $N$ order and select the one which gives correlation greater than a correlation threshold (defined in next section). Once coefficients are extracted, we then create a linear filter.

The next step is to generate a white Gaussian noise $Z$ with mean ($\mu$) of 0 and different noise scale levels. The random noise $Z$ is passed through a linear filter to generate a correlated noise $Z'$. The correlated noise is then added to the forecasted data $X$ using Eq. \ref{noise_add}. \vspace{-1.5mm}
\begin{equation}
    X' = X +Z'
    \label{noise_add}
\end{equation}

The noisy correlated data $X^{'}$ is then released to remote app servers. Algorithm \ref{algo:2} illustrates the algorithmic description of the privacy preservation phase. 
\begin{algorithm}[!t]
\caption{Time-Series Privacy Preservation}
\DontPrintSemicolon
\SetKwInOut{Input}{Input}\SetKwInOut{Output}{Output}
  \Input{$X$ = real-time time-series data of user $u$, \newline $V$ = pre-trained forecasting models belonging to $u$, \newline $C$ = pre-trained clustering models belonging to $u$, \newline $F$ = set of time-series where each represents a feature in $X$, \newline $\tau$ = correlation threshold, $N$ = order}
  \Output{$X'$= obfuscated $X$}
\For{$f\leftarrow 1$ \KwTo $F$}{
Input $f$ into $C$, select a cluster label $l$ from $C$ and the corresponding forecasting model $v$ from $V$ \\
Determine the stability of $f$ using ADF, apply Box-Cox transformation to unstable $f$ \\
Forecast $f$ using $v$ to get forecasted time-series $f'$ \\
Perform reverse Box-Cox transformation on $f$' using Eq.(\ref{eq:inv-box}) \\
\For{$n\leftarrow 1$ \KwTo $N$}{
$R = -1$

Create a linear filter $L$ \\
\While{$R < \tau$}{
    Generate white Gaussian Noise $Z$ \\
    Pass $Z$ through $L$ to get correlated noise $Z'$ \\
    Calculate Correlation $R$ between $X'$ and $Z'$ \\
}
Add $Z'$ to $f'$ using Eq. (\ref{noise_add})\\
$X' = X' \cup f'$
}
}
\label{algo:2}
\end{algorithm}
\section{Experiment Settings} 
\label{sec:eval_mob}
In this section, we first briefly describe the datasets, evaluation metrics, and experimental procedure used in our experiments.

\subsection{Datasets}
We use three datasets to demonstrate the effectiveness of our obfuscation mechanism. We use (a) Pen-based Gestures ~\cite{beuvens2012usigesture} and (b) Mobile Swipes~\cite{masood2018touch} as our experimental datasets. These two datasets demonstrate scenarios where a user is writing on a touchscreen or swiping on a touchscreen, respectively. All these tasks are performed by users either using fingers or stylus pen in different sessions. For readability, we use handwriting and swipes to  refer these datasets in the rest of the paper. Table~\ref{tab:stats1} shows the summary statistics of these datasets. The values are shown for total number of users, samples in each dataset, different functionality types (e.g. alphabets, swipe direction), and raw features associated with each dataset. We collected a total of 603,933 samples from a total of 595 users across all the datasets. 
\begin{table}[!b]
\centering
\tiny
\captionsetup{skip=0.3pt}
  \caption{\tiny Statistics of Datasets}
  \label{tab:stats1}
  \begin{tabular}{|c|c|c|c|c|}
    \hline
     \textbf{Dataset} & \textbf{\# of} & \textbf{\# of} & \textbf{Possible} & \textbf{\# of Raw}\\
      & \textbf{Users} & \textbf{Samples} & \textbf{Types} & \textbf{Features}\\
      \hline
    HW(L) & 30 & 7,800 & 26 & 5\\ \hline
    HW(D) & 30 & 3,000 & 10 & 5\\ \hline
    Swipes & 35 & 129,100 & 4 & 3 \\ \hline
Total: & 95 & 139,900 & & \\ \hline
\end{tabular}
\end{table}

\textbf{Handwriting Dataset:} To demonstrate the applicability of our obfuscation mechanism in a latest and user-centered technology such as stylus, we use a dataset provided by Beuvens et al.~\cite{beuvens2012usigesture}. The original purpose of collecting this dataset was to help developers in determining the most suitable gesture recognition algorithm in various contexts/scenarios. The features used in this dataset could be used in the context of mobile app where a user uses stylus pen to write on his mobile device to perform a certain functionality. In our experiments, we use gesture features for 26 letters of alphabets and 10 numeric digits from a total of 30 participants. Each alphabet and digit is recorded 10 times making a total of 300 gestures per letter/numeric digit. We apply our obfuscation mechanism separately on letters and digits so as to validate its effectiveness on different writing formats. We refer \textit{HW(L)} to letters dataset and \textit{HW(D)} to digits dataset in the rest of the paper.

\textbf{Swipes Dataset:} This dataset contains touch swipes performed by users on mobile devices for various purposes such as scrolling documents, images, and playing games. The original purpose of the dataset~\cite{mahbub2016active} was to investigate the automated user authentication techniques on the mobile devices. We filtered out all the swipes which have less than 5 data points, and finally extracted 129,100 samples from a total of 35 users from this dataset.

\textbf{Features:} We use raw features from the datasets to evaluate the efficacy of our method. We use four features from the Handwiring dataset i.e. \textit{x-position, y-position, pen orientation, pen angle, and pen pressure}. For Swipes dataset, we use three raw features, which are \textit{x-position, y-position, and finger pressure}.

\subsection{Evaluation Metrics}
\label{metric}
In this section, we define privacy and utility metrics to measure the performance of our obfuscation mechanism. 

\textbf{Untrackability:} We define "untrackability" as an inability of an adversary or a third-party to track a user across different sessions. For example, the gesture data of a user reading document at two different time, should not exhibit any similarity that makes an adversary identifies the user. Instead, the adversary should think that the two gesture instance are performed by two different users. Theoretically, let $X^{'}_{I_1}$ denote the obfuscated time-series data of a user $u$ at a time interval $I_1$ consisting of $n$ data points $\{x^{'}_{1,1}, x^{'}_{1,2}, ....x^{'}_{1,n}\}$ whereas $X^{'}_{2}$ is another obfuscated time-series of the same user taken at another time interval $I_2$, i.e. $\{x^{'}_{2, 1}, x^{'}_{2, 2}, ....x^{'}_{2, n}\}$. We then define ``untrackability'' as an inability to link both $X^{'}_{1}$ and $X^{'}_{2}$ to a user $u$ based on the distance or similarity between $X^{'}_{1}$ and $X^{'}_{2}$.  

 For our work, we use the Mean Absolute Error (MAE) as a distance metric to measure untrackability of a user across sessions. MAE is a common measure of forecast error in time series analysis \cite{armstrong1992error}. Given two series $X^{'}_{1}$ and $X^{'}_{2}$ of a user $u$, MAE is calculated as:\vspace{-2.3mm}
\begin{equation}
    \text{Untrackability} = \frac{1}{n} \sum_{i=1}^n  |x'_{1,i} - x'_{2, i}|
\end{equation}
i.e. our obfuscation mechanism should have high MAE between different time-series data of the same user indicating that a user cannot be tracked across sessions. 

\textbf{Indistinguishability:} This metric identifies whether an adversary is able to uniquely identify or distinguish a user's data from a set of other users' data. 
Assume that $D$ is a dataset containing $n$ time-series, indistinguishability is the likelihood where $k$ time-series of $D$ are $X'_t$, $k \leq n$, 
based on the similarity or distance between $X^{'}_{t}$ of $u$ and time-series data of $(k-1)$ users is: \vspace{-2.3mm}
\begin{equation*}
    H(X^{'}_{t}|D) \geq \log_2{k}
\end{equation*}
and the Information Gain (IG) is:\vspace*{-2mm}
\begin{equation*}
    IG(X^{'}_{t}|D) \leq \log_2{n} - \log_2{k}
\end{equation*}
\vspace{-3.5mm}
\begin{equation}
    IG(X^{'}_{t}|D ) \leq \log_2 {n/k}
\end{equation}
Thus, distinguishability indicates how much information an adversary can infer about a user to distinguish him/her from other users and indistinguishability is the inverse of it. The higher the $IG(X^{'}_{t}|D)$ is, the lower is the indistinguishability. 

\textbf{Utility:} The utility refers to correctly performing the intended functionality of an app, for instance, correctly recognizing letters (such as a, b, c) and swipe directions (left, right, up, and down) from the obfuscated data. In our experiments, we measure utility by computing MAE between original time-series $X$ and obfuscated time-series $X'$ as given in Eq. \ref{eq:ut}. \vspace{-2.3mm} 
\begin{equation}
    \text{Utility} = \frac{1}{n} \sum _{i=1}^{n}\left|x_i -x'_{i}\right|,
    \label{eq:ut}
\end{equation}
where $X$ is an original time-series with data points $\{x_1, x_2, ....x_n\}$ and $X'$ is an obfuscated time-series having $\{x'_{1}, x'_{2}, ....x'_{n}\}$ data points. A high value for MAE indicates low utility and vice versa.

\textbf{Remark 2:} Readers might confuse between the difference between the untrackability and utility metrics as both are measured using MAE. The difference is based on the data taken by both metrics for calculations. In case of untrackability, obfuscated time-series at different time intervals are taken such that error rate the error rate indicates whether the user is trackable or not. Utility metric, on the other hand, takes into account the error between obfuscated time-series $X^{'}_i$ and original time-series $X_i$ such that high error rate indicates that the obfuscated time-series is highly different from the original time-series (i.e. low utility) and vice versa. 

\subsection{Experimental Setup}
\label{setup}

\textbf{Data Pre-processing:} 

First, the data is filtered by removing  broken, invalid, or empty values. Each time-series data, i.e. a gesture, in the datasets is labelled with the numbers such as $1, 2,...$ to indicate its sequence and start and end. Next, we fix sampling rate of each gesture to 90 percentile of data points. This is necessary because mobile devices send data at a fixed sampling rate. After fixing sampling rate, we normalize datasets using min-max scaling \cite{aksoy2001feature}. This step is performed to make illustration of results consistent across different datasets. Finally, we partitioned the data from each dataset into two sets. The first set has 80\% of the samples, and the second has the remaining 20\% of the samples. The larger set – 80\% was used as training dataset, while 20\% of the samples were used for ``testing''. The training samples are used to train cluster and forecasting models, whereas testing samples are used to check the effectiveness of our obfuscation mechanism at run-time. 

\begin{figure*}[!t]
\tabcolsep=0.11cm
\captionsetup{skip=0pt, justification=centering}
\captionsetup[subfloat]{farskip=2pt,captionskip=1pt, font=scriptsize}
\centering
\subfloat[HW(L)]{\label{ns_HW}
      \includegraphics[scale=0.09, keepaspectratio]{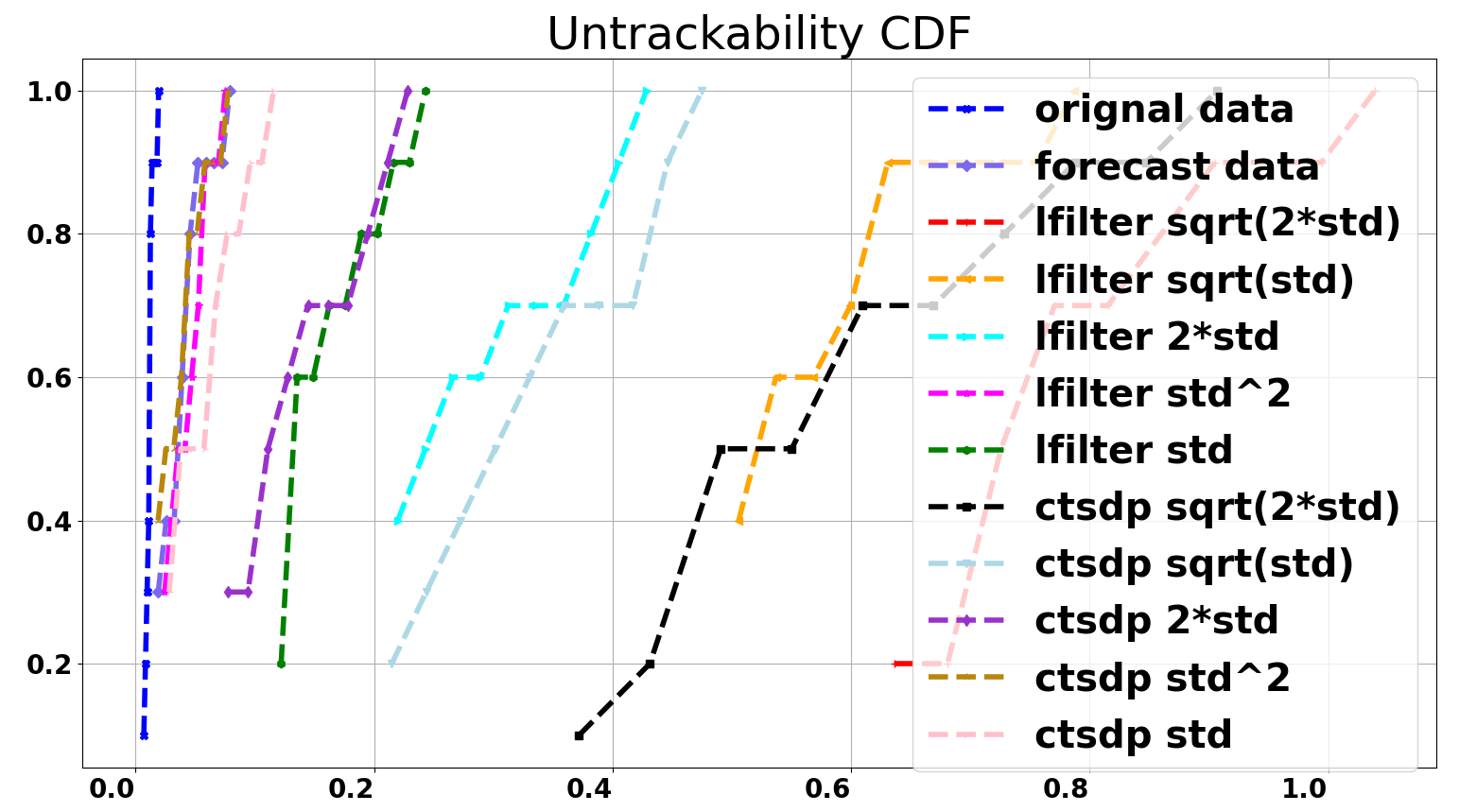}}
\subfloat[HW(D)]{\label{ns_Dig}
      \includegraphics[scale=0.09, keepaspectratio]{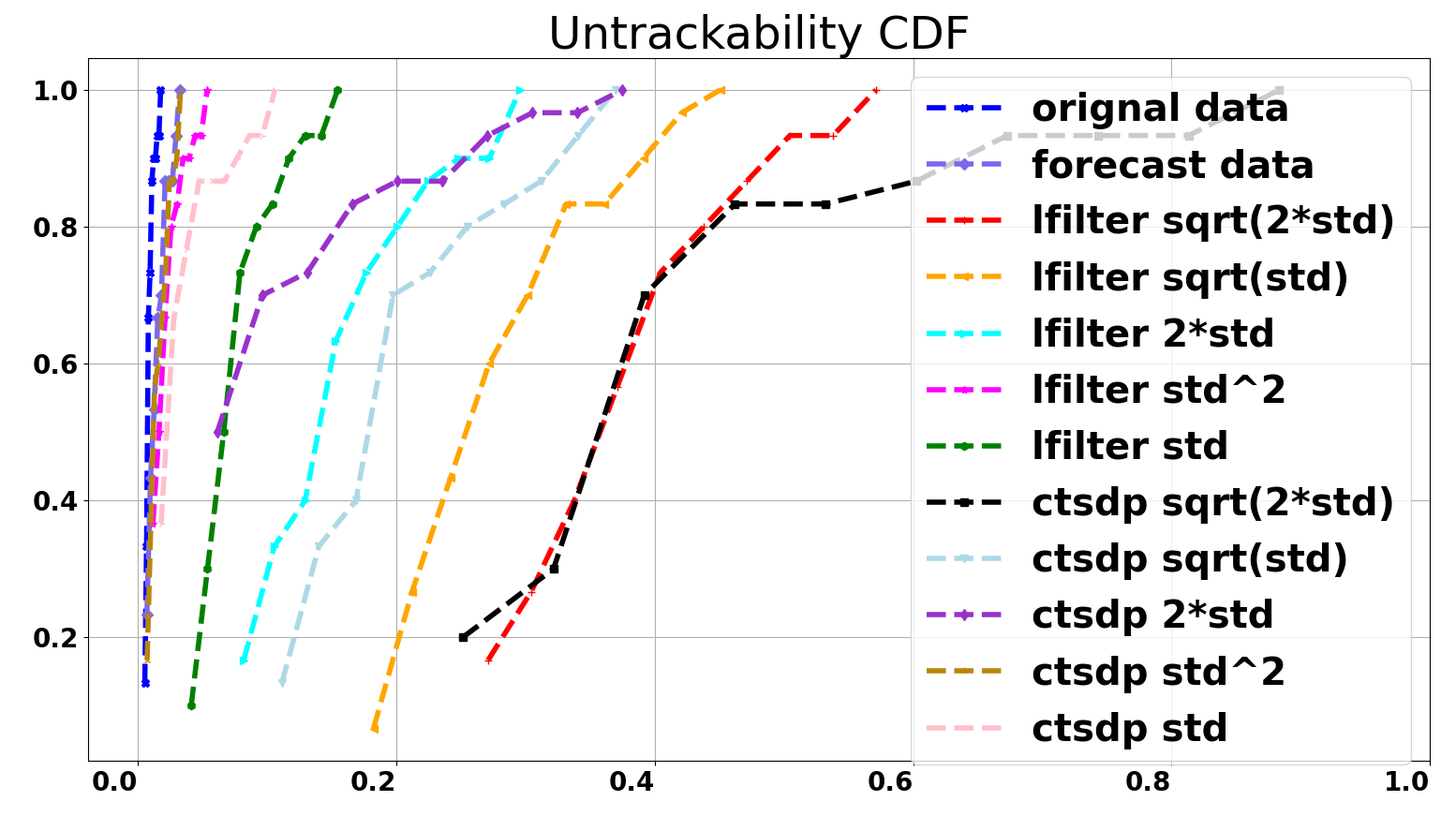}}
\subfloat[Swipes]{\label{ns_Swipes}
      \includegraphics[scale=0.115, keepaspectratio]{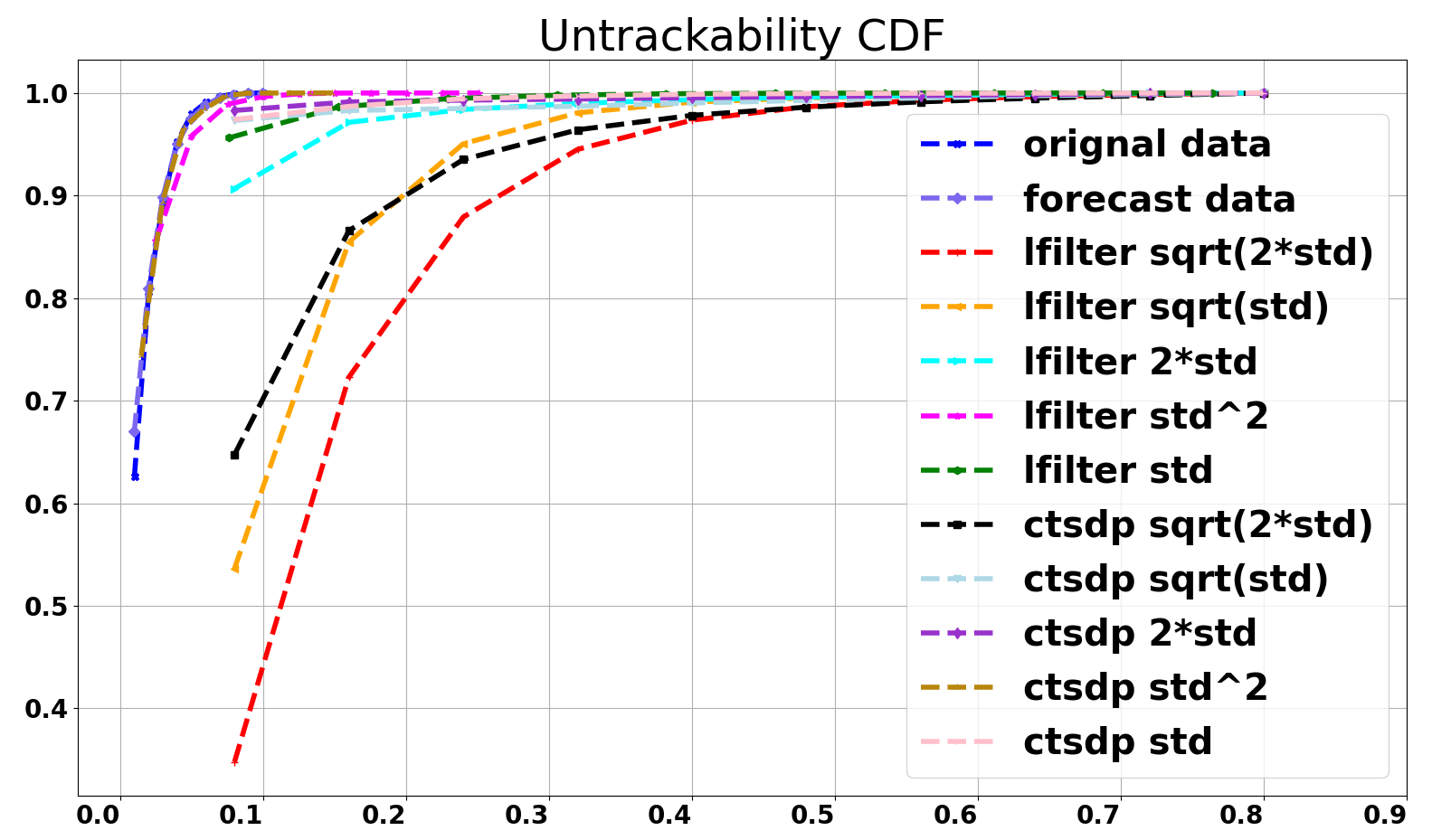}}
\caption{CDF of untrackability at varying Guassian noise scale levels in all datasets. Y-axes represent fraction of the participant population and X-axes show untrackability, respectively.}
\label{cdf_untrack}
\end{figure*}
\begin{figure*}[!t]
\tabcolsep=0.11cm
\captionsetup{skip=0pt, justification=centering}
\captionsetup[subfloat]{farskip=2pt,captionskip=1pt, font=scriptsize}
\centering
\subfloat[HW(L)] {\label{cs_HW}
       \includegraphics[scale=0.09, keepaspectratio]{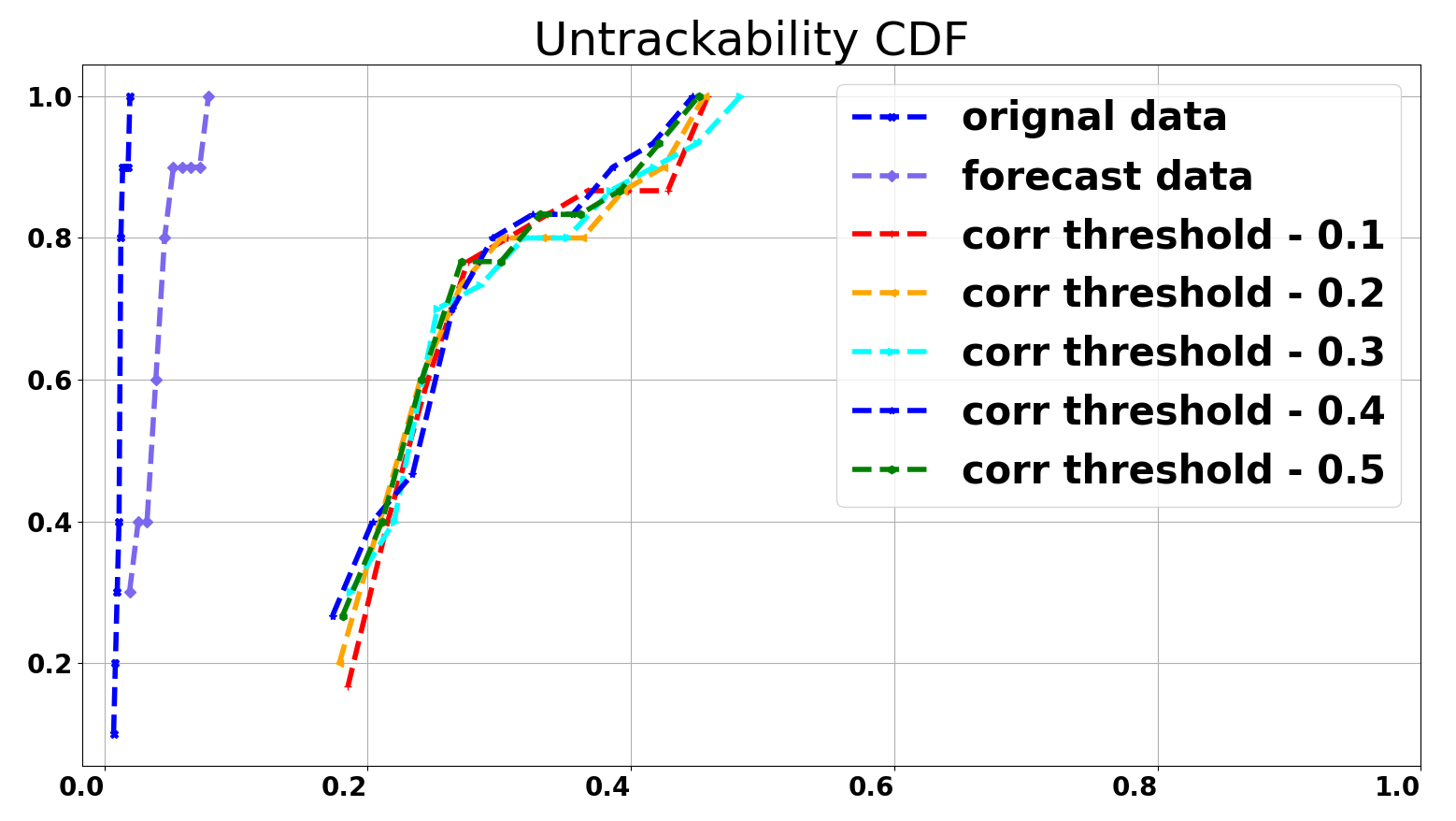}}
\subfloat[HW(D)] {\label{cs_Dig}
       \includegraphics[scale=0.09, keepaspectratio]{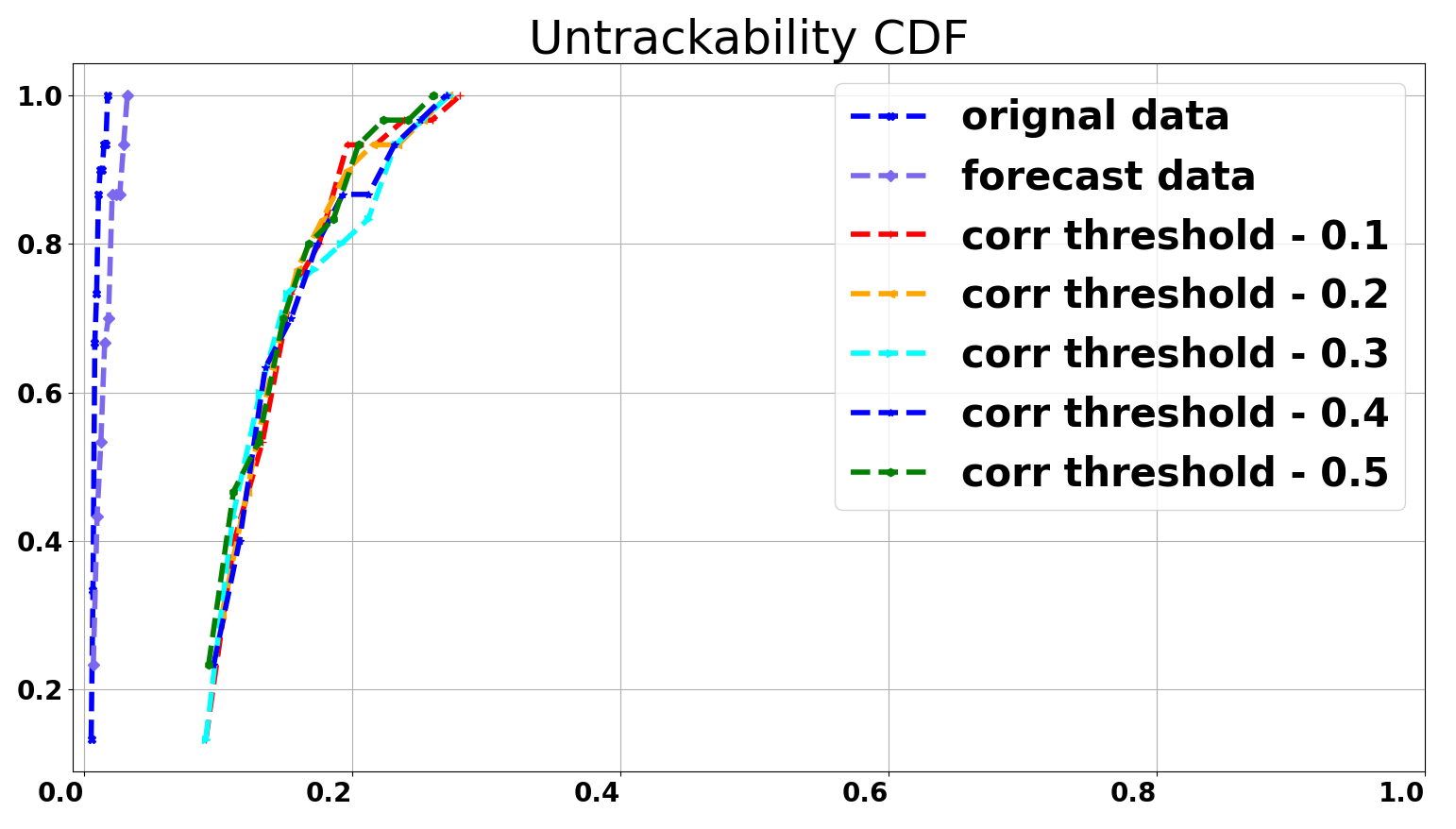}}
\subfloat[Swipes] {\label{cs_Swipes}
       \includegraphics[scale=0.108, keepaspectratio]{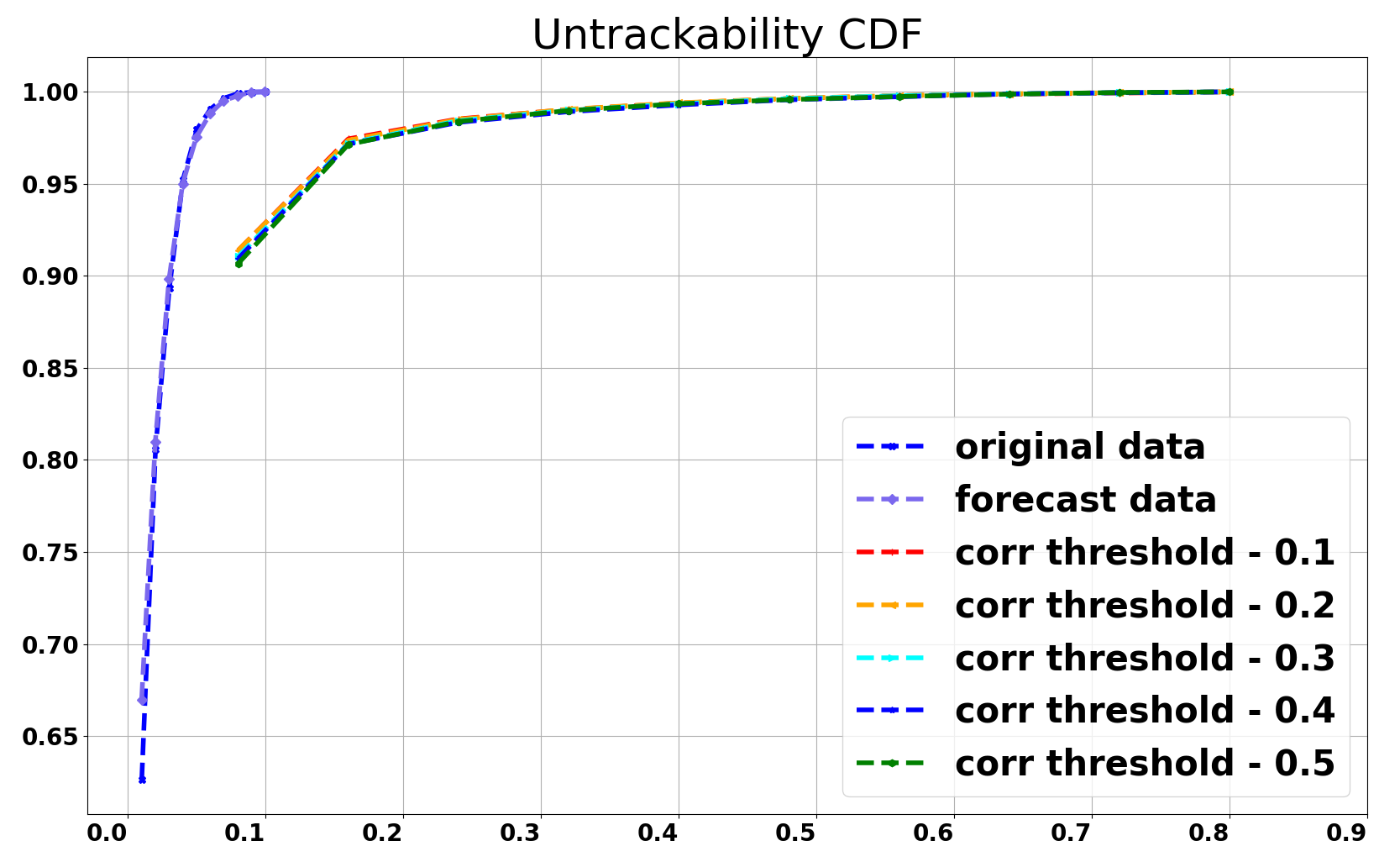}}
\caption{CDF of untrackability at varying correlation threshold values in all datasets. Y-axes represent fraction of the participant population and X-axes show Untrackability, respectively.}
\label{cdf-corr_untrack}
\end{figure*}

\textbf{Parameter Setting:} For each dataset, the obfuscation mechanism is evaluated against ``Gaussian Noise Scale'', and ``Correlation Threshold'' parameters, where the former is tested for 5 different values [sqrt(2*std), sqrt(std, 2*std, std\textasciicircum 2, std)], and the latter is tested for 5 different values ranging from 0.1 to 0.5 with 0.1 interval. The ``Gaussian Noise Scale'' parameter generates random noise within the given scale whereas, the ``Correlation Threshold'' parameter decides the correlation level of noise. The different values of these two parameters explain the trade-off between privacy and utility i.e. an increase in a noise scale indicates high privacy but less utility, whereas a decrease in noise scale indicates good utility but less privacy. On the contrary, an increase in correlation threshold reflects low privacy but high utility.

\section{Evaluation}
\label{res_sec}

In this section, we specifically evaluate our proposed mechanism on how well it (a) prevent the tracking and distinguishing a user, and (b) preserves the app utility after obfuscation. We run experiments in two settings: In the first setting, we choose ``Gaussian Noise Scale'' to be one of: square root of 2 times standard deviation of the incoming series (sqrt(2*std)), square of root standard deviation of the incoming series (sqrt(std)), 2 times the standard deviation (2*std), standard deviation squared (std\textasciicircum 2) and standard deviation (std), with fixed ``Correlation Threshold`` of 0.5. In the second setting, we fixed ``Gaussian Noise Scale'' to be 2 times standard deviation (2*std) and  change ``Correlation Threshold'' from 0.1 to 0.5 with an increment of 0.1.

\subsection{Untrackability}
\label{sec:untrack}

We first measure untrackabilty of original datasets which shows a lower MAE between different sessions of the same user. A lower MAE value (i.e. low untrackability) implies that the two time-series from two different sessions of the same user are quite similar and thus has a higher risk of being tracked by an adversary. Our results show that (1) in the original datasets users can be tracked across different sessions based on their gestures to perform certain activities e.g. write or swipe. We found (2) similar results for forecasted data since it is a stabilized version of the original data. On the other hand, (3) untrackability is significantly improved after applying our obfuscation mechanism on the datasets.

\begin{figure*}[!b]
\tabcolsep=0.11cm
\captionsetup{skip=0pt, justification=centering}
\captionsetup[subfloat]{farskip=2pt,captionskip=1pt, font=scriptsize}
\centering
\subfloat[HW(L)] {\label{fig:noise_IG}
      \includegraphics[scale=0.09, keepaspectratio]{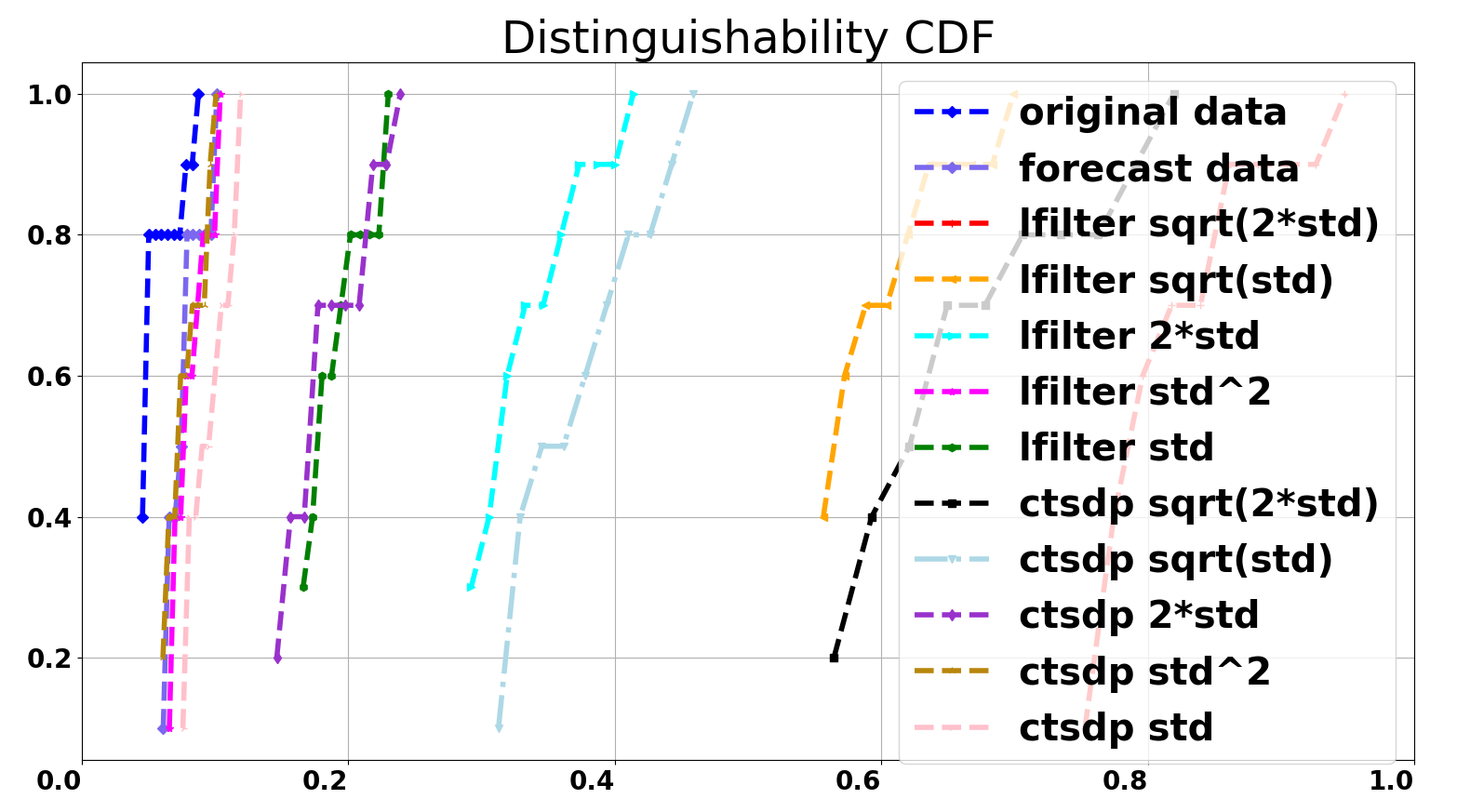}}
\subfloat[HW(D)] {\label{fig:Dig_IG}
      \includegraphics[scale=0.09, keepaspectratio]{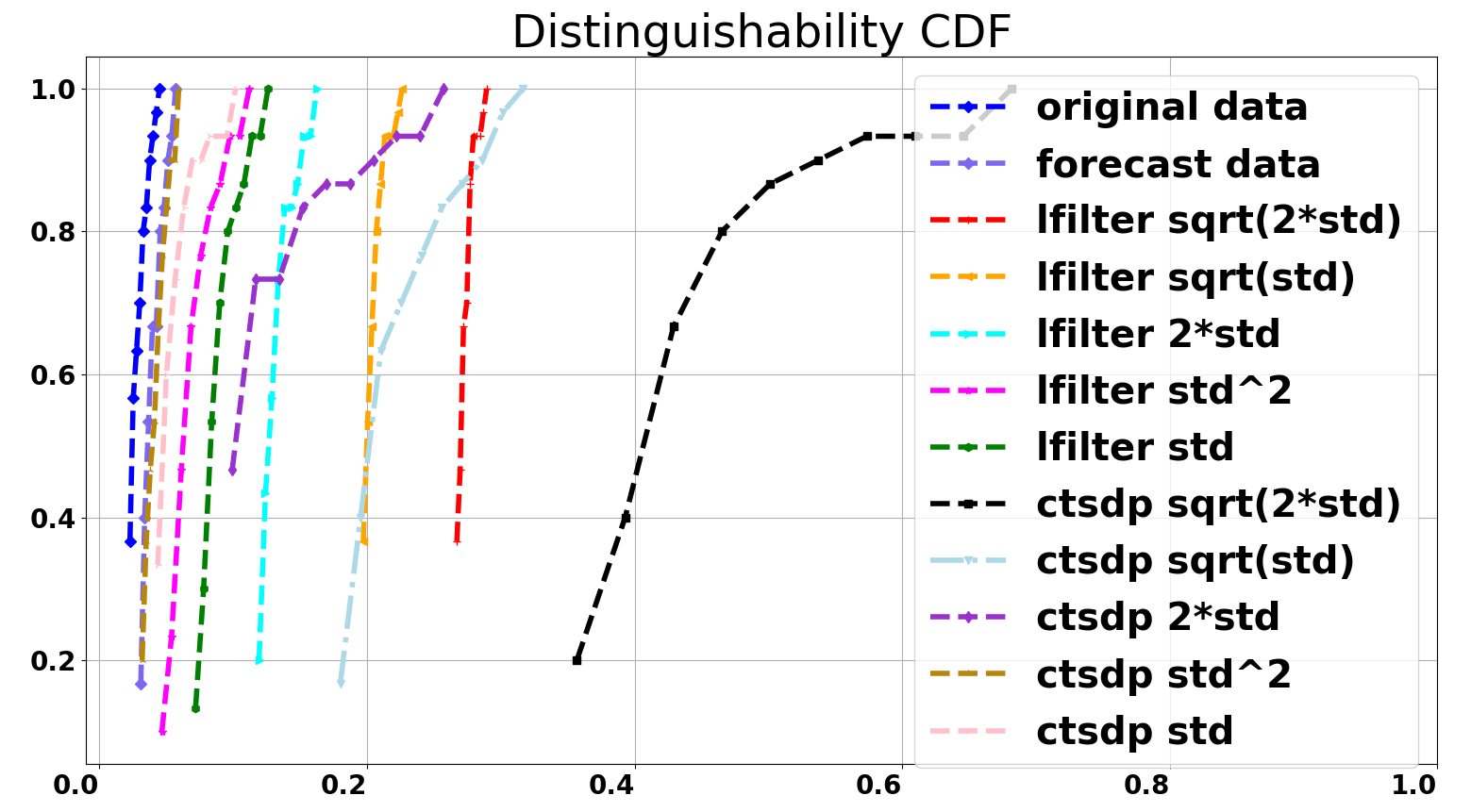}}
\subfloat[Swipes] {\label{fig:Swipes_IG}
      \includegraphics[scale=0.114, keepaspectratio]{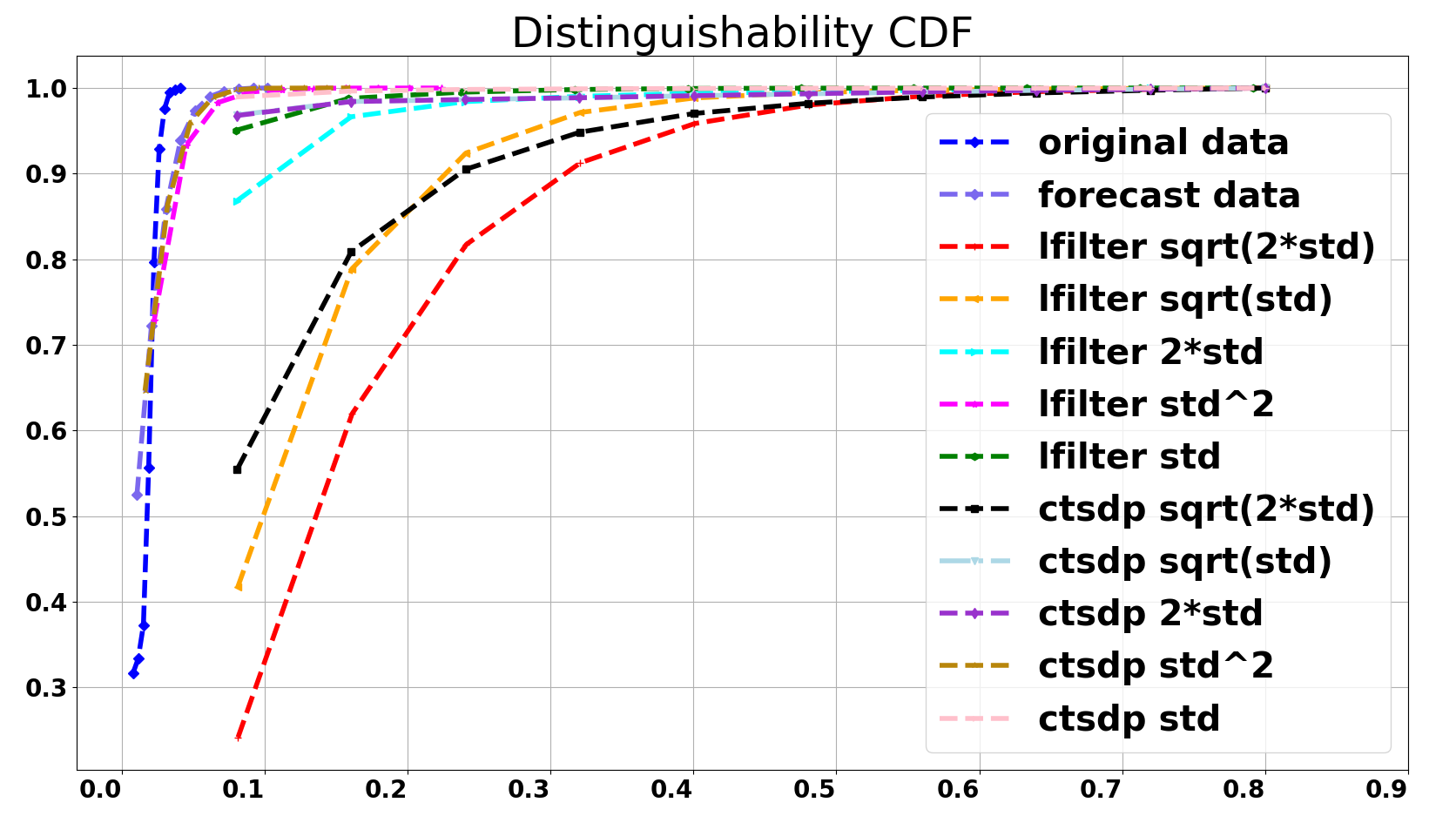}}
\caption{CDF of indistinguishability at varying Guassian noise scale levels in all datasets. Y-axes represent fraction of the participant population and X-axes show indistinguishability, respectively.}
\label{cdf-indis}
\end{figure*}
\begin{figure*}[!b]
\tabcolsep=0.11cm
\captionsetup{skip=0pt, justification=centering}
\captionsetup[subfloat]{farskip=2pt,captionskip=1pt, font=scriptsize}
\centering
\subfloat[HW(L)] {\label{fig:corr_IG}
      \includegraphics[scale=0.09, keepaspectratio]{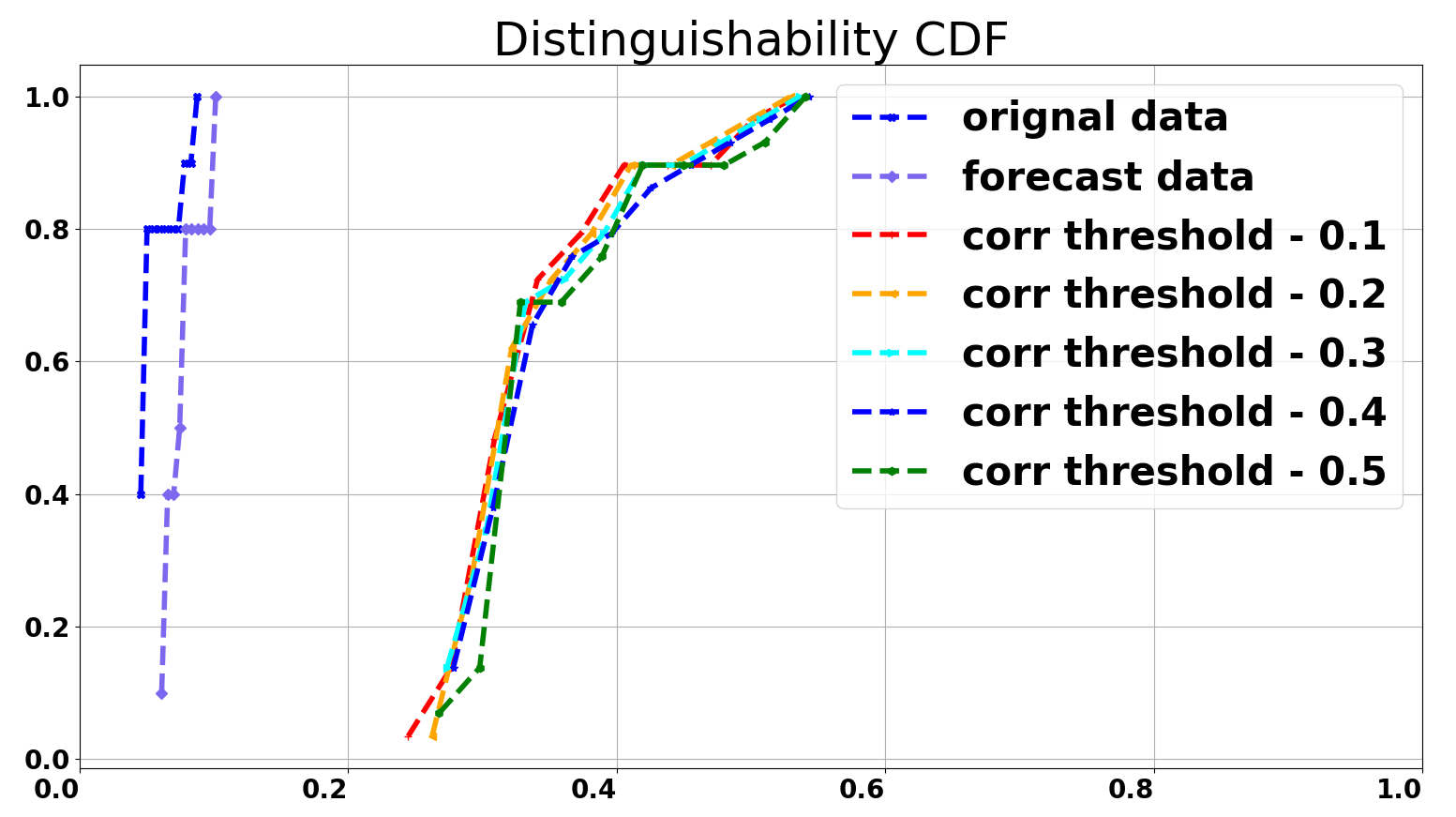}}
\subfloat[HW(D)] {\label{fig:corr_IG_Dig}
      \includegraphics[scale=0.09, keepaspectratio]{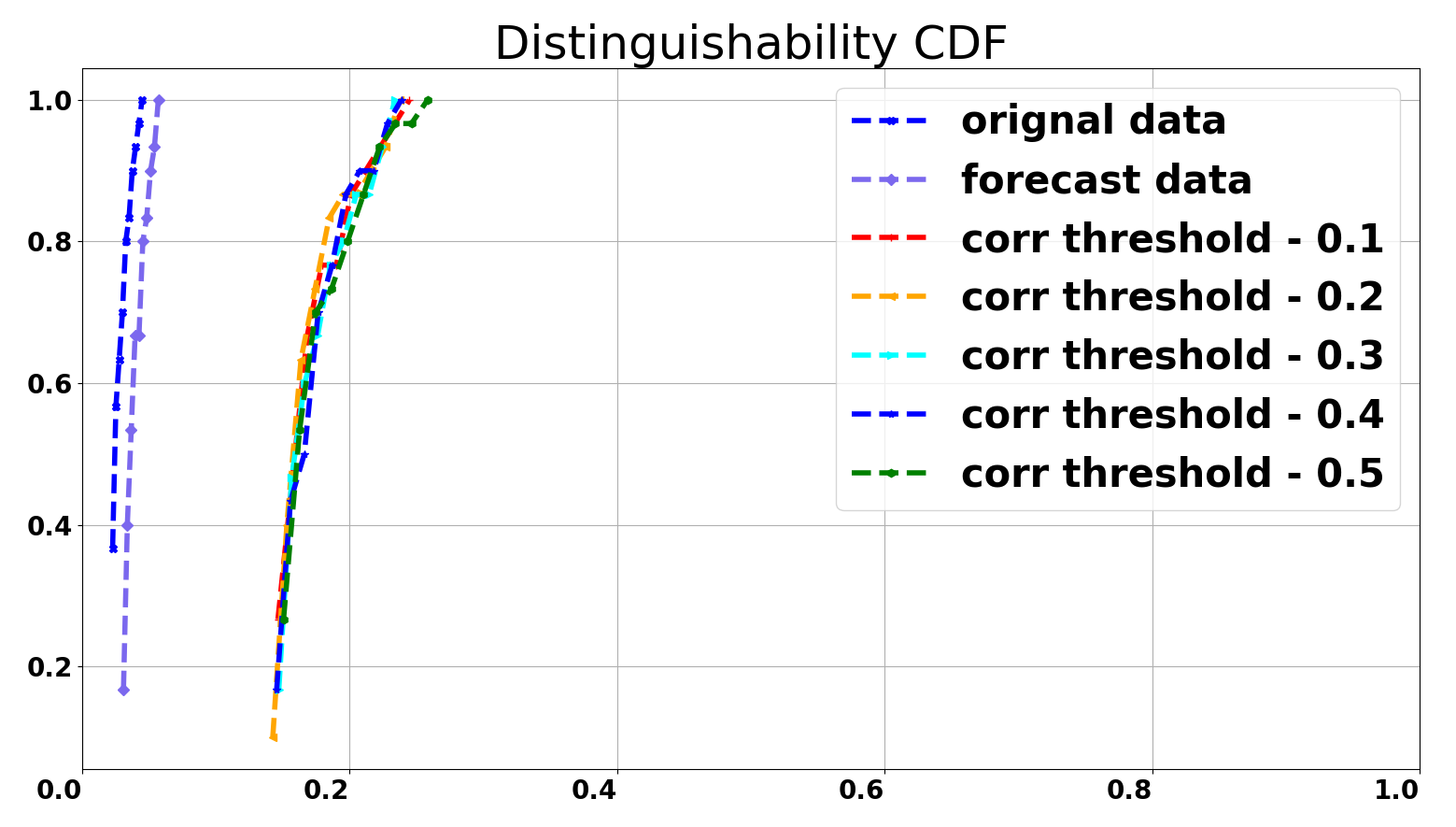}}
\subfloat[Swipes] {\label{fig:corr_IG_Swipes}
      \includegraphics[scale=0.112, keepaspectratio]{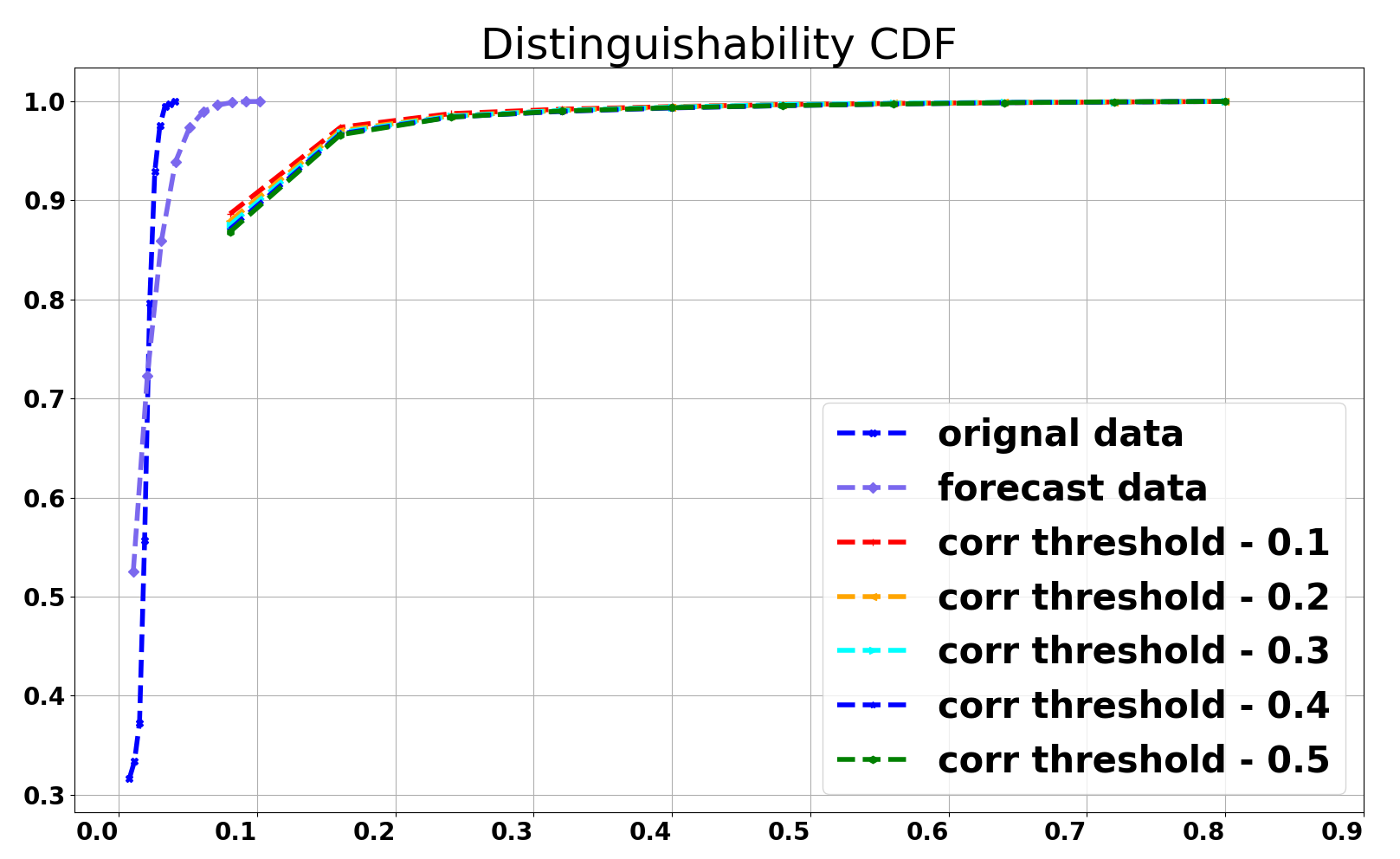}}
\caption{CDF of indistinguishability at varying correlation threshold values in all datasets. Y-axes represent fraction of the participant population and X-axes show indistinguishability, respectively.}
\label{cdf-corr_indis}
\end{figure*}

Fig.~\ref{cdf_untrack} shows Cumulative Distribution Function (CDF) of untrackability in the original, forecasted, and obfuscated data against 5 noise scales. For the HW(L) dataset (Fig.~\ref{ns_HW}), assuming that all the users initially have 0\% untrackability, their untrackability increases to 9\% with the forecasted data and then eventually increases to 25\% with additional noise generated using our proposed linear filter with the noise scale being the standard deviation of the forecasted series. Our analysis on HW(D) dataset shows that original and forecasted data have 3\% and only 4\% of untrackability, which reaches to 45\% for 80\% of users in the obfuscated data using proposed linear filter with noise scale being sqrt(2*std) (Fig.~\ref{ns_Dig}).
We observe that untrackability is highly improved with the Swipes dataset for example, with all noise scales, 80\% of users have untrackability of nearly 100\% (Fig.~\ref{ns_Swipes}). Hence, the results from Swipes datasets indicate that the high number of samples positively impacts the results of our obfuscation mechanism.

We also examine results with different correlation threshold values ranging from 0.1 to 0.5, and a fixed noise scale of 2 times the standard deviation of the incoming series (2*std). Results are shown in Fig.~\ref{cdf-corr_untrack}, indicating zero correlation between untrackability and correlation threshold, i.e. increasing correlation between noise and the original data will not change untrackability across sessions. For HW(L) dataset, the untrackability is about 24\% for 50\% of users with all correlation thresholds (Fig.~\ref{cs_HW}). In Fig. \ref{cs_Dig}, we show the untrackability of HW(D) dataset where maximum untrackability rate is close to 20\% for 80\% of users. 
Similarly, in Swipes dataset and for all correlation thresholds, untrackability reaches to 80\% for all users. Figure \ref{cs_Swipes} shows results from Swipes dataset.

\begin{figure*}[!b]
\tabcolsep=0.11cm
\captionsetup{skip=0pt, justification=centering}
\captionsetup[subfloat]{farskip=2pt,captionskip=1pt, font=scriptsize}
\centering
\subfloat[HW(L)]{\label{ns_ut_HW}
      \includegraphics[scale=0.09,
      keepaspectratio]{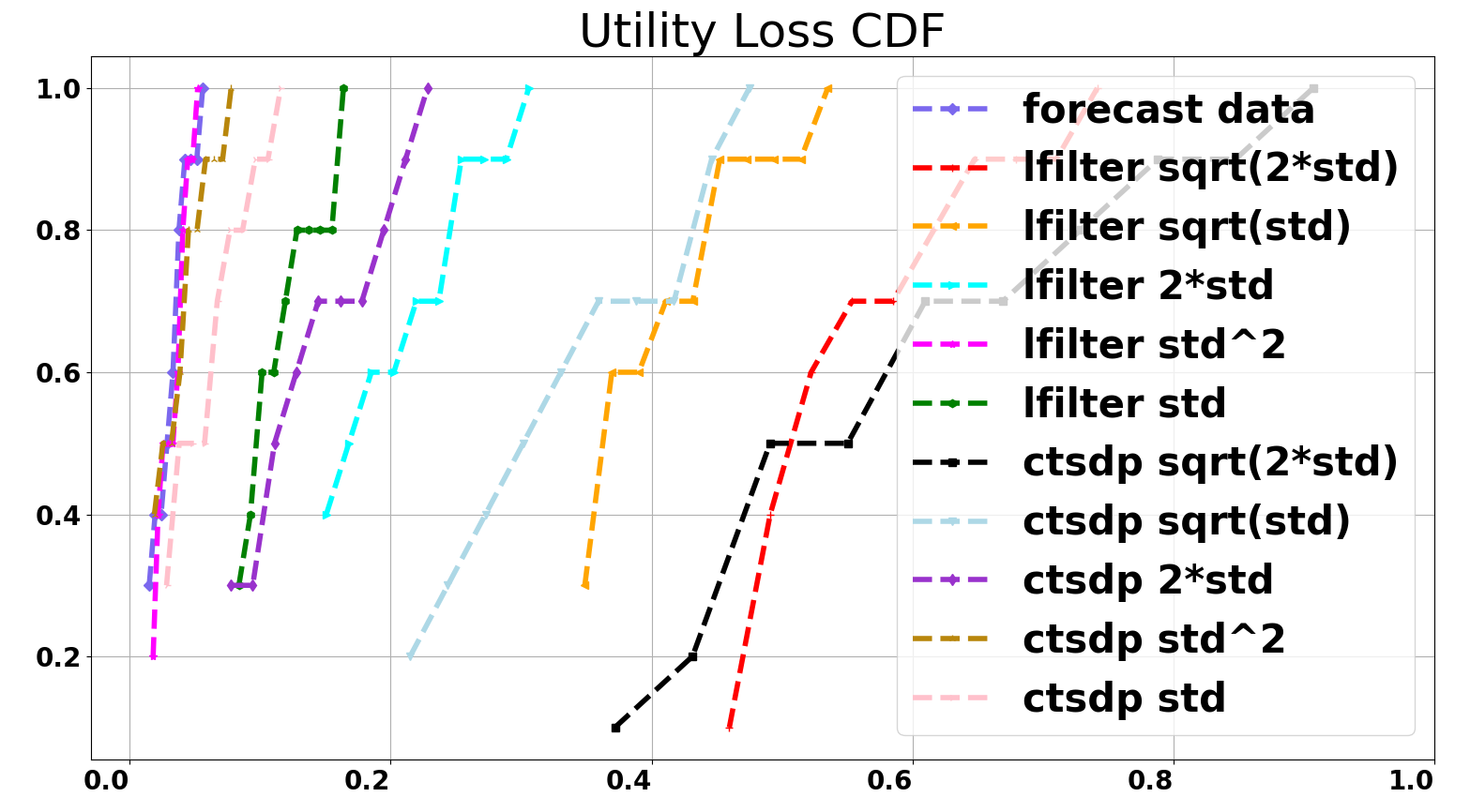}}
\subfloat[HW(D)]{\label{ns_ut_Dig}
      \includegraphics[scale=0.09,
      keepaspectratio]{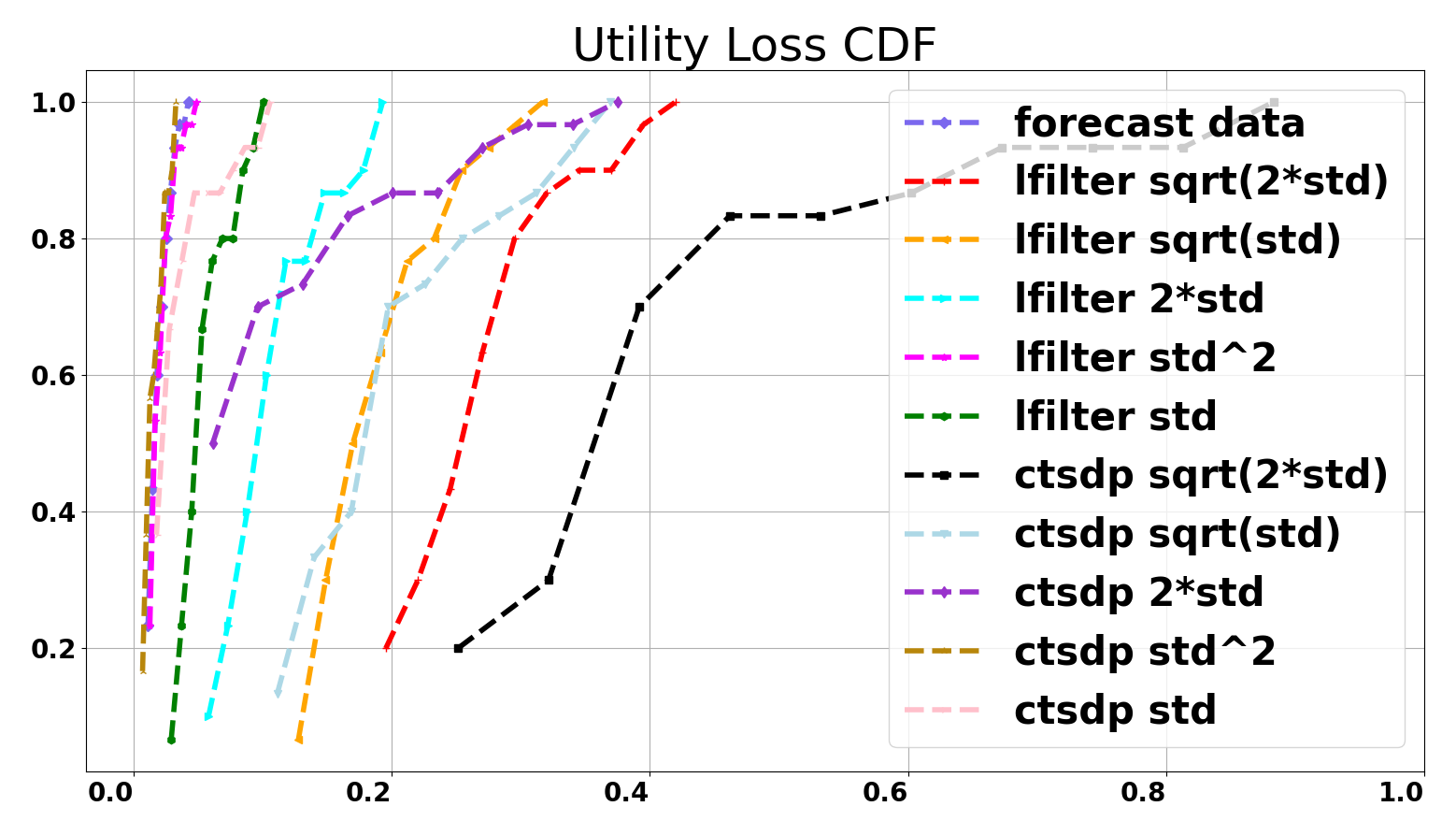}}
\subfloat[Swipes]{\label{ns_ut_Swipes}
      \includegraphics[scale=0.116,
      keepaspectratio]{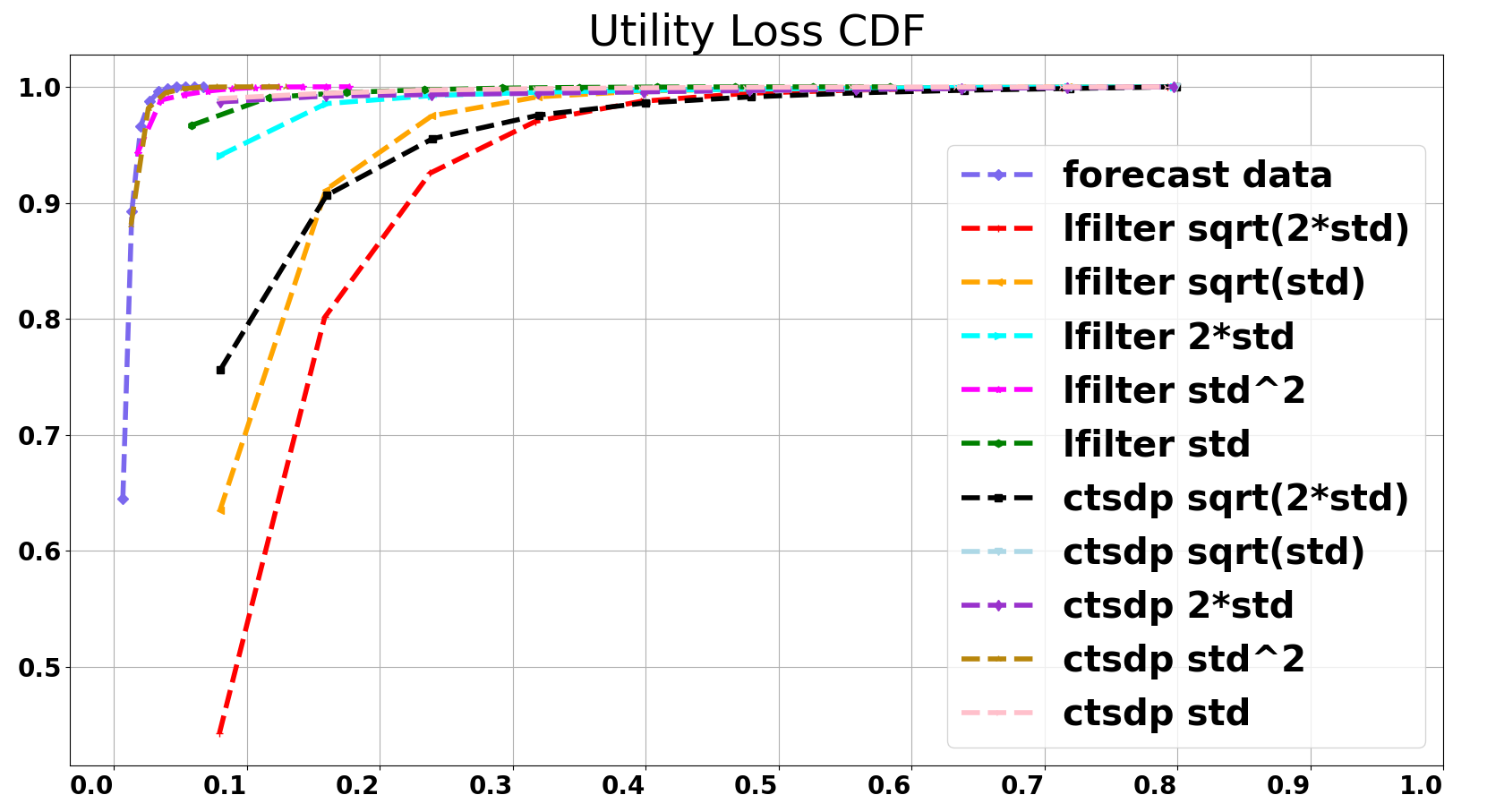}} 
\caption{CDF of Utility Loss at varying Guassian noise scale levels in all datasets. Y-axes represent fraction of the participant population and X-axes show Utility (MAE), respectively.}
\label{cdf-Utility}
\end{figure*}
\begin{figure*}[!b]
\tabcolsep=0.11cm
\captionsetup{skip=0pt, justification=centering}
\captionsetup[subfloat]{farskip=2pt,captionskip=1pt, font=scriptsize}
\centering
\subfloat[HW(L)] {\label{cs_ut_HW}
      \includegraphics[scale=0.09,
      keepaspectratio]{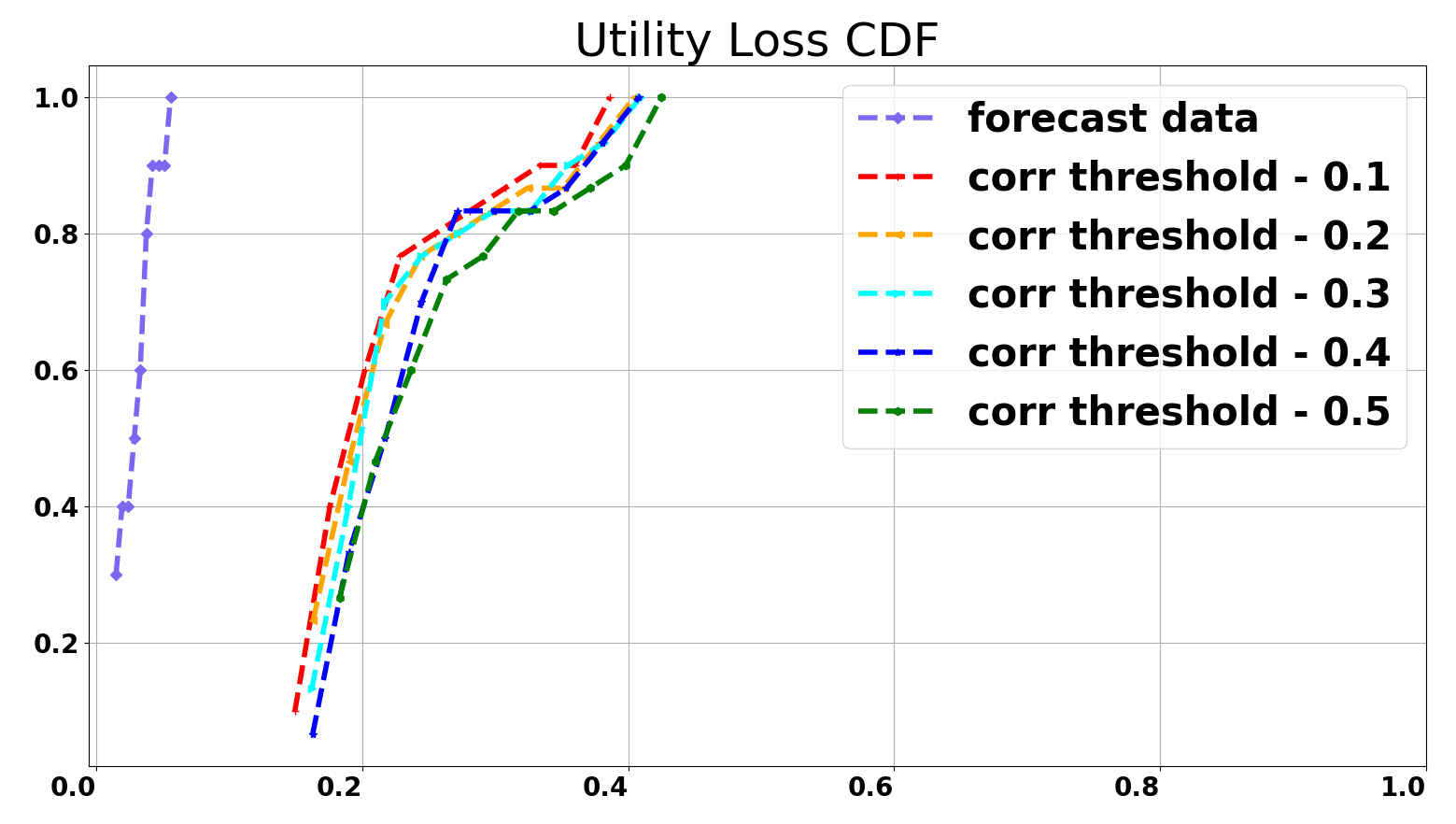}}
\subfloat[HW(D)] {\label{cs_ut_Dig}
      \includegraphics[scale=0.09,
      keepaspectratio]{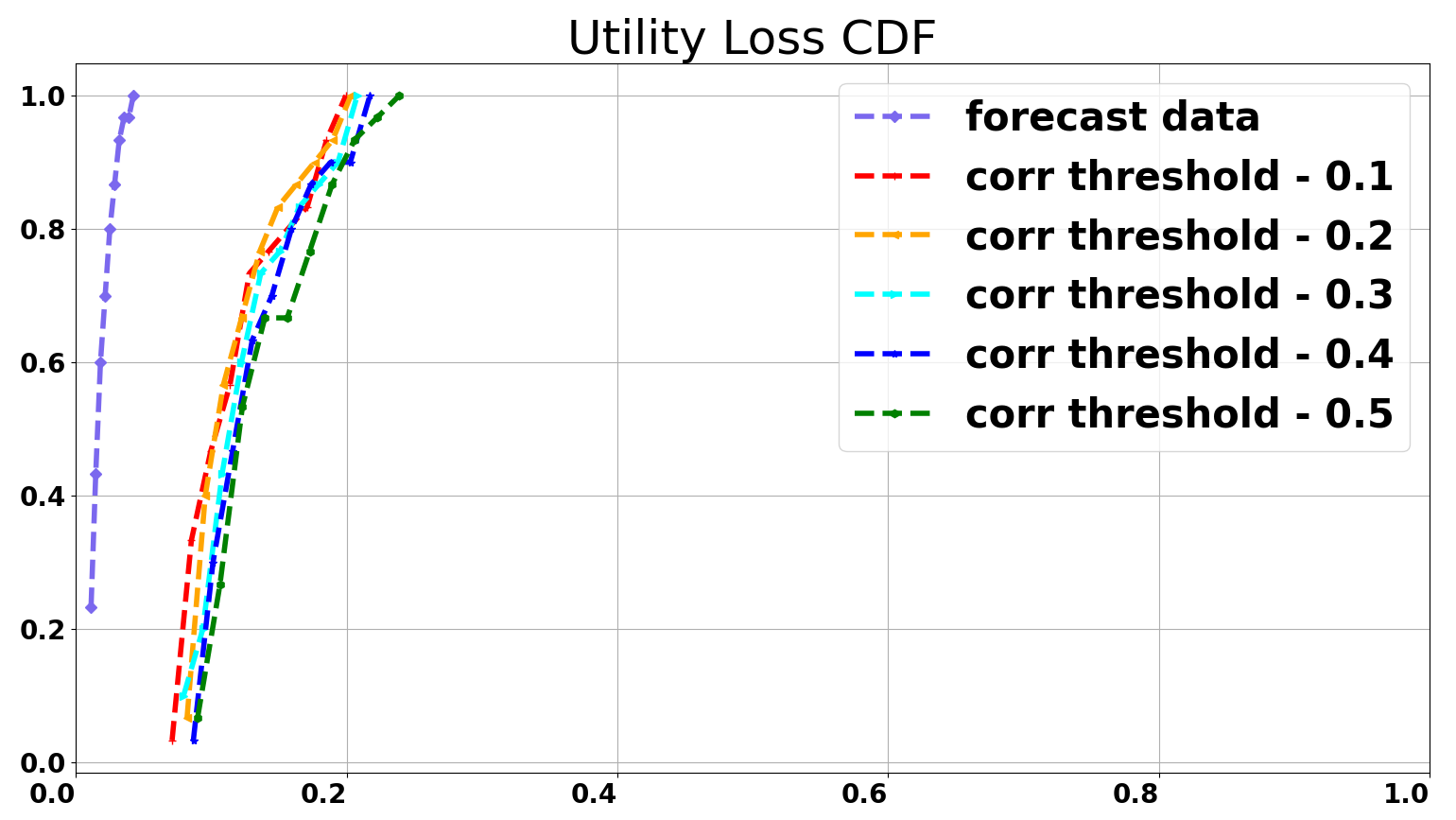}}
\subfloat[Swipes] {\label{cs_ut_Swipes}
      \includegraphics[scale=0.102,
      keepaspectratio]{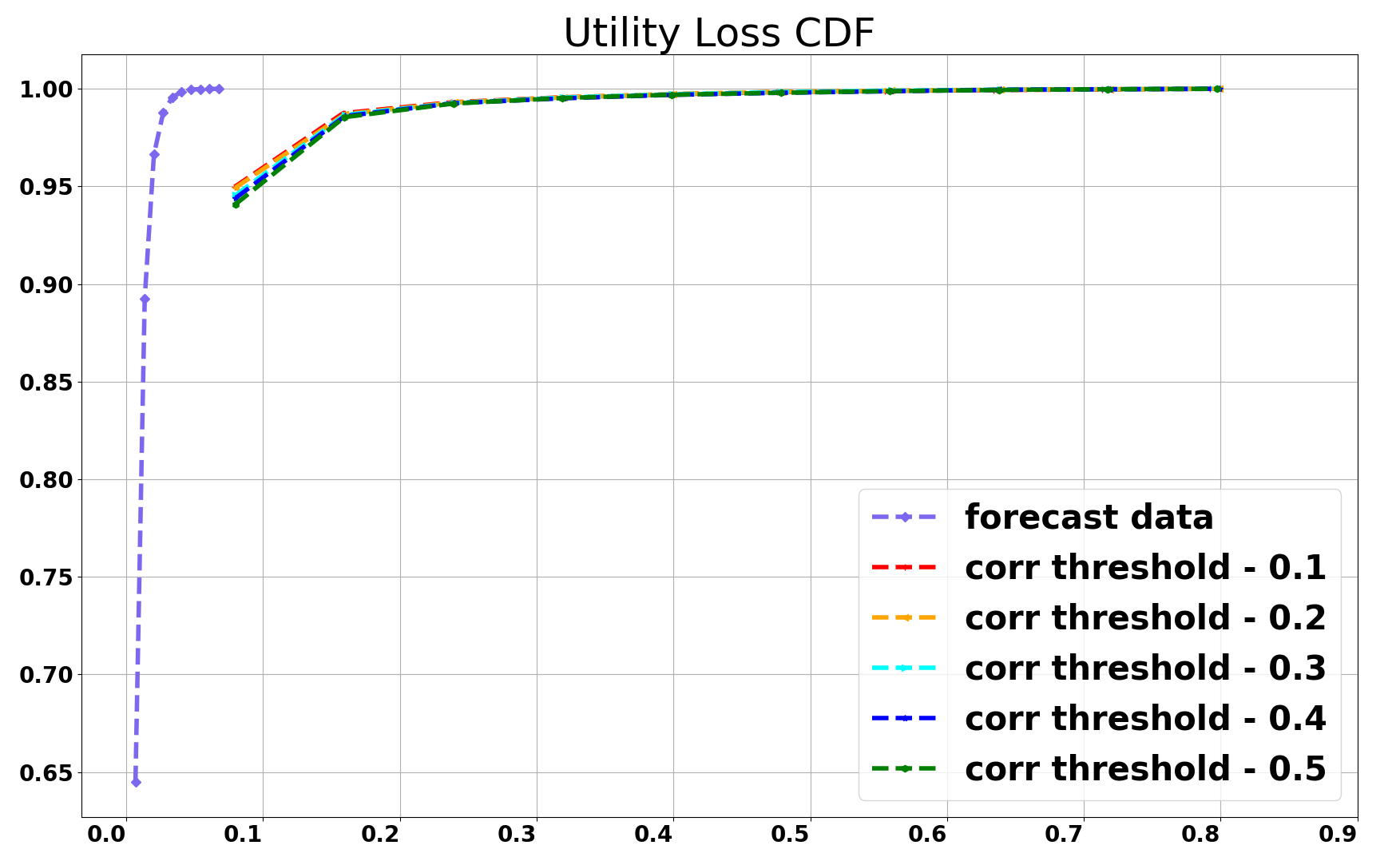}}
\caption{CDF of Utility Loss at varying correlation threshold values in all datasets. Y-axes represent fraction of the participant population and X-axes show Utility (MAE), respectively.}
\label{cdf-corr_utility}
\end{figure*}

In general, it is clearly evident from the results of all datasets that increasing the noise scale increases the untrackability, thus increasing privacy. The untrackability can be further increased if we fix the noise scale to be square root of 2 times the standard deviation.  However, to balance between the utility and privacy, appropriate values for these two parameters need to be set.

\subsection{Indistinguishability}
\label{res_indis}
Next, we evaluate our mechanism using the distinguishability metric. We found that an increase in noise scale increases distinguishability. Fig.~\ref{cdf-indis} and~\ref{cdf-corr_indis} show CDFs of distinguishability against different noise scale and correlation values, respectively. For the HW(L) dataset (Fig.~\ref{fig:noise_IG}), we found that the distinguishability increases to 23\% for all users, with a noise scale being the standard deviation of the incoming series. For HW(D) dataset, the mechanism gives 11\% of distinguishability for all users when noise scale is the standard deviation (Fig.~\ref{fig:Dig_IG}). 
The results on the Swipes dataset shows that indistinguishability increases significantly with the noise scale being 2*std. For instance, with a noise scale of std, 10\% of users have reached indistinguishability close to 95\% (Fig.~\ref{fig:Swipes_IG}).

We observe similar trend across different correlation threshold values as shown in Fig.~\ref{cdf-corr_indis}, where the distinguishability is consistent with varying correlation threshold. For HW(L) dataset (Fig.~\ref{fig:corr_IG}), 50\% of users achieve 30\% of distinguishability. HW(D) dataset offers distinguishability of 17\% for 60\% of users (Fig.~\ref{fig:corr_IG_Dig}). 
Similarly, any change to correlation threshold does not necessarily change the distinguishability of users. (Fig.~\ref{fig:corr_IG_Swipes}).

\subsection{Utility} 
\label{utility_res}

Our results on preserving the functionality (utility) of an app show that reasonable level of utility could be achieved by tuning the parameter values (Fig.~\ref{cdf-Utility} and ~\ref{cdf-corr_utility}). The utility requirement in HW(L) dataset is to correctly recognize letters that are written from the stylus-pen.  We observe that forecasted data has a utility as low as 0.3. However, after the obfuscation, the MAE starts increasing with the increase in noise scale value. With the noise scale of std\textasciicircum 2, the MAE is 0.2 for 50\% of users and 0.3 for 80\% of users. We observe that increasing noise scale to sqrt(2*std) results in MAE of 0.5 for 50\% of users (Fig.~\ref{ns_ut_HW}).
The MAE of HW(D) dataset is 0.27 for 60\% of users with the noise scale of sqrt(2*std) which eventually decreases with the decrease in noise scale values (Fig.~\ref{ns_ut_Dig}). 
Our results on the utility loss of Swipes dataset show a loss of only 16\% for a noise scale of sqrt(2*std) for 80\% of the users, which decreases as noise scale changes to sqrt(std) (fig.~\ref{ns_ut_Swipes}). For instance, 80\% of users have utility loss of only 13\% with a noise scale of sqrt(std).

Fig.~\ref{cs_ut_HW} shows the effect of obfuscation mechanism on the HW(L) dataset with different correlation threshold values. The increase in correlation threshold will not increase or decrease utility loss but not as close to the forecasted data. The reason perhaps is that obfuscation mechanism is bounded with a fixed noise scale value. Similarly, Fig.~\ref{cs_ut_Dig} illustrates the MAE of HW(D) dataset for different correlation threshold values. We observe that with all correlation thresholds, MAE is 0.12 for 60\% of users which increases to 0.2 for all user. 
For the Swipes dataset, utility loss for 95\% of users with a correlation scale of 2*std (Fig.~\ref{cs_ut_Swipes}).

\begin{figure*}[!t]
\tabcolsep=0.11cm
\centering
\subfloat[]{\label{feat_untrack_HW}
      \includegraphics[scale=0.09, keepaspectratio]{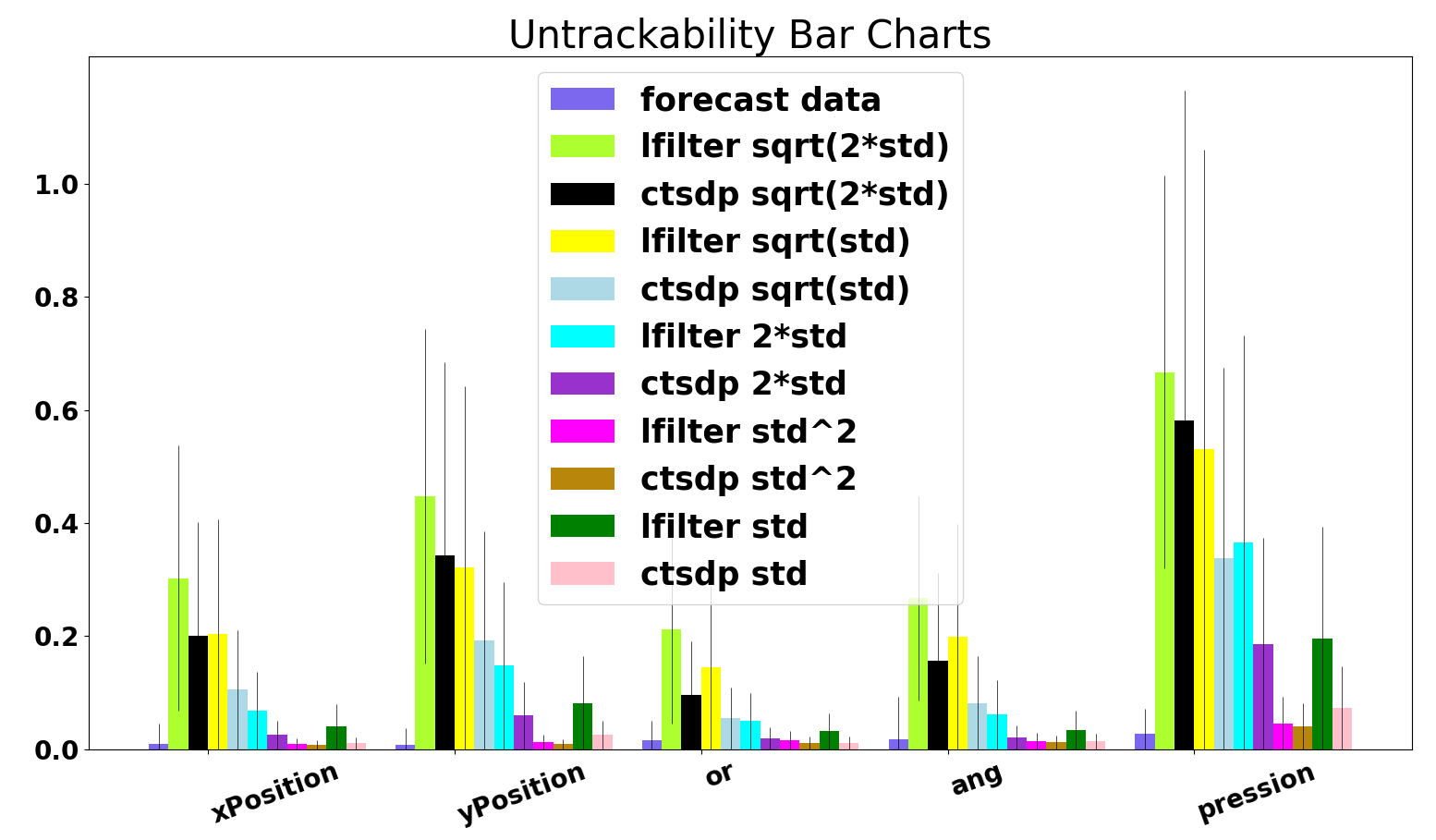}}
\subfloat[]{\label{feat_untrack_Dig}
      \includegraphics[scale=0.09,  keepaspectratio]{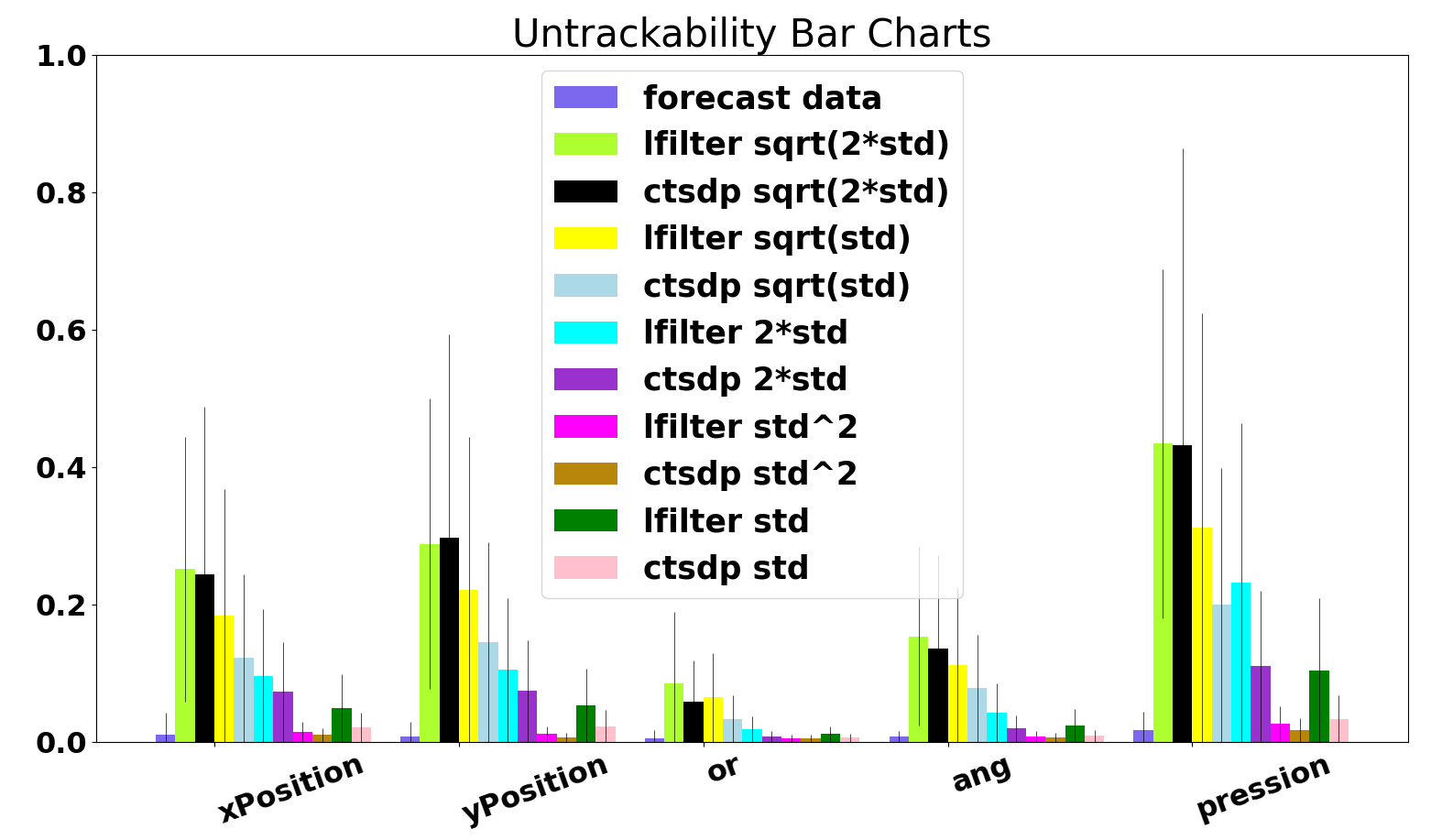}}
\subfloat[]{\label{feat_untrack_Swipes}
      \includegraphics[scale=0.099, keepaspectratio]{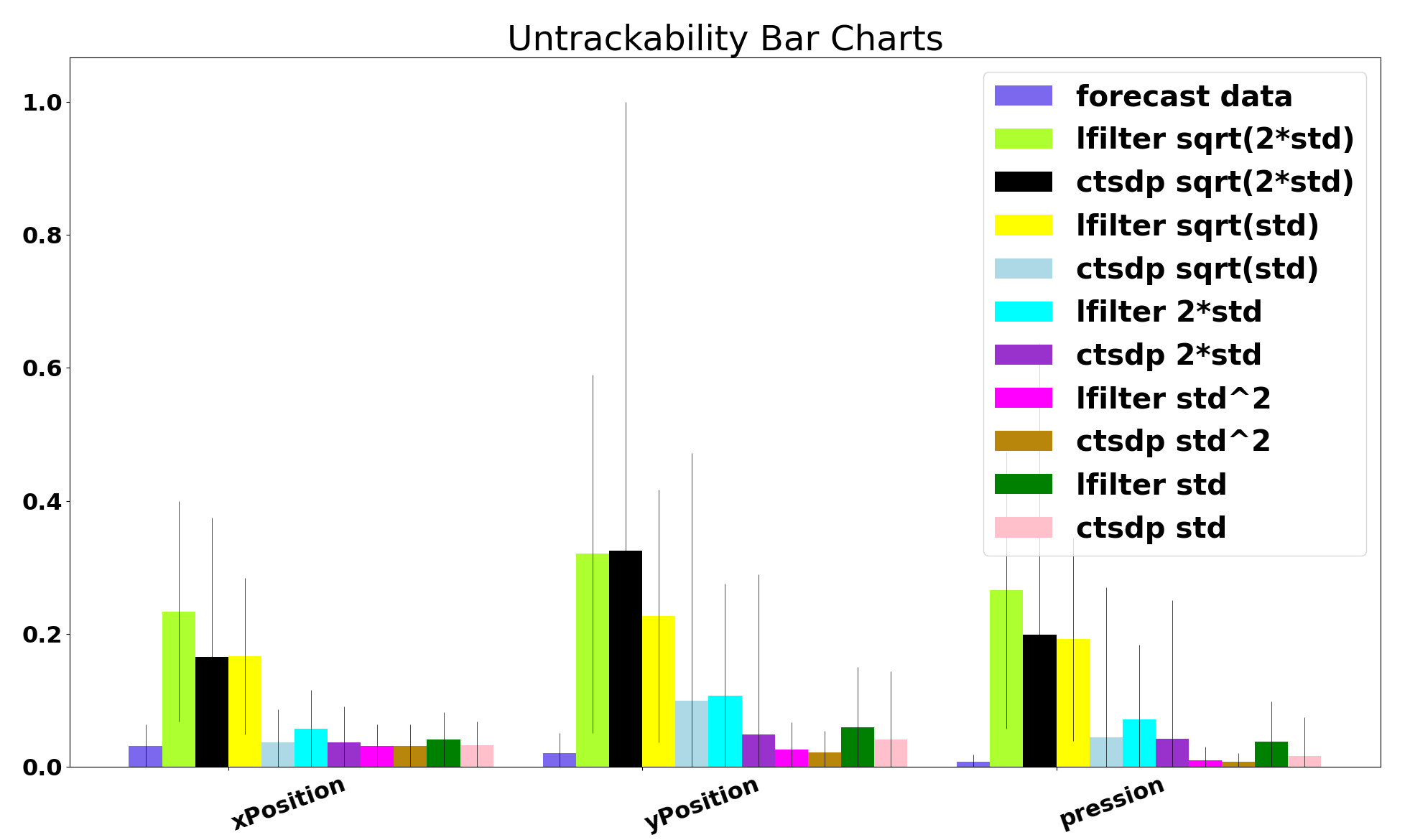}}
\end{figure*}
\begin{figure*}[!t]
\tabcolsep=0.11cm
\centering
\subfloat[]{\label{feat_IG_HW}
      \includegraphics[scale=0.09, keepaspectratio]{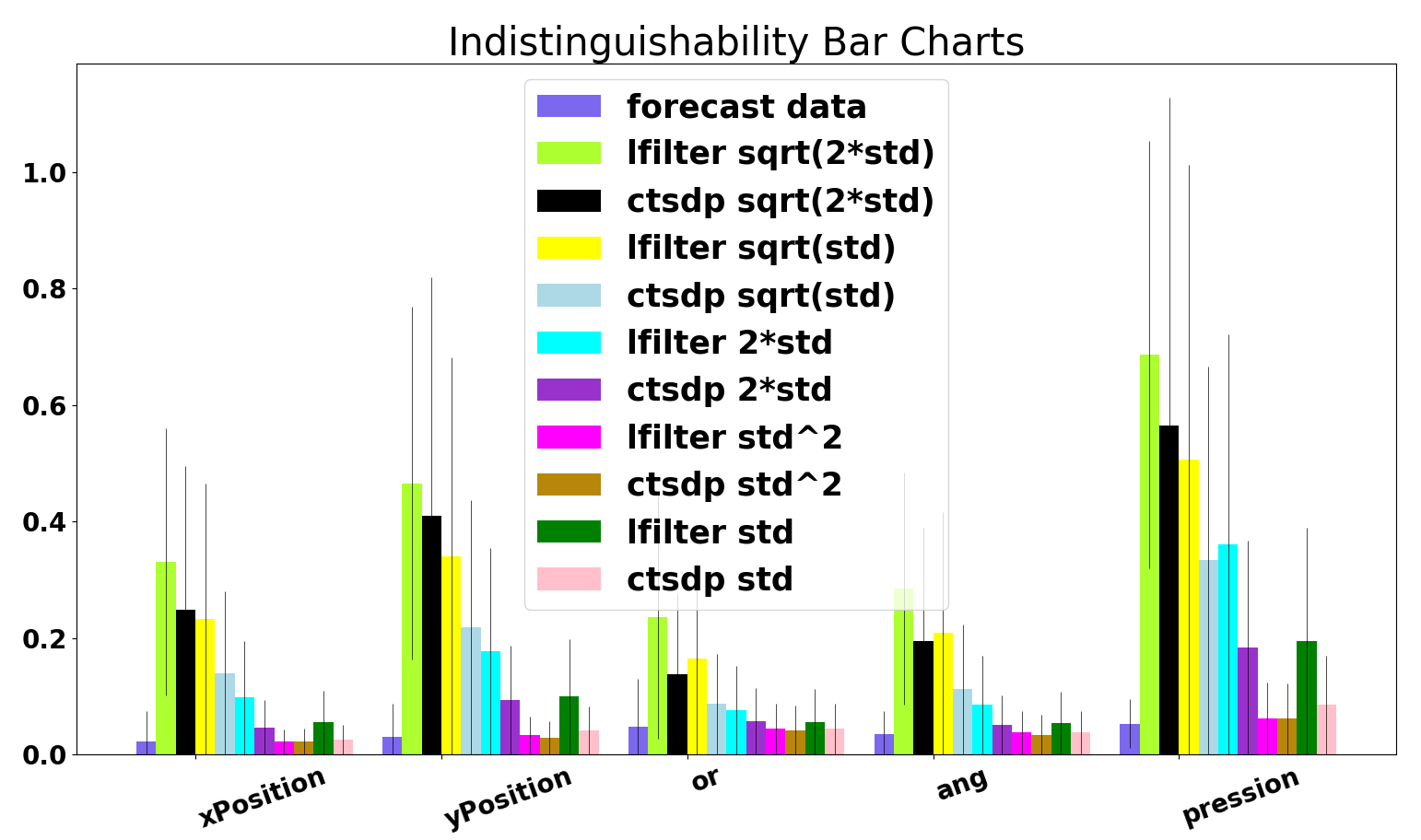}}
\subfloat[]{\label{feat_IG_Dig}
      \includegraphics[scale=0.09, keepaspectratio]{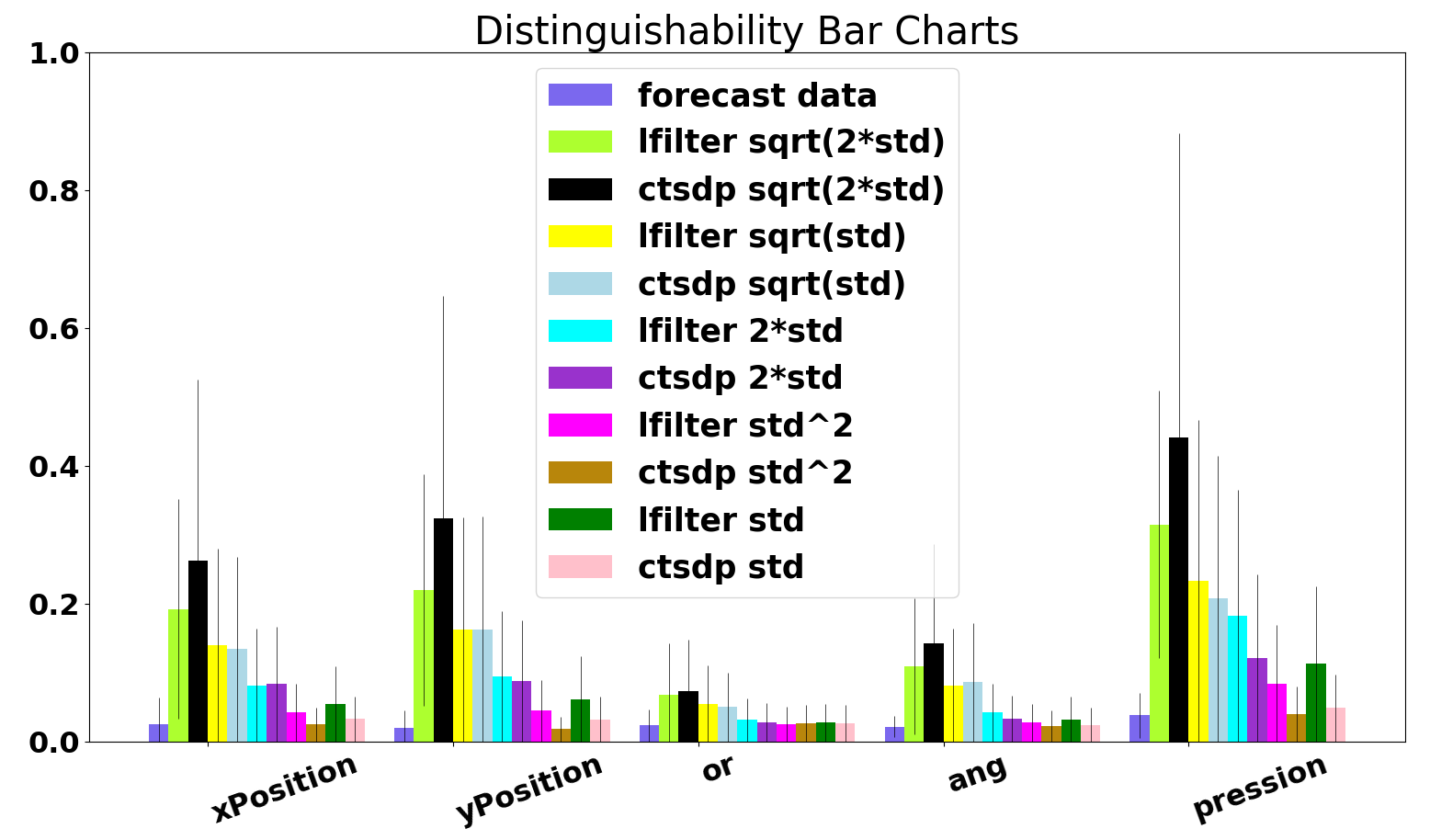}}  
\subfloat[]{\label{feat_IG_Swipes}
      \includegraphics[scale=0.105, keepaspectratio]{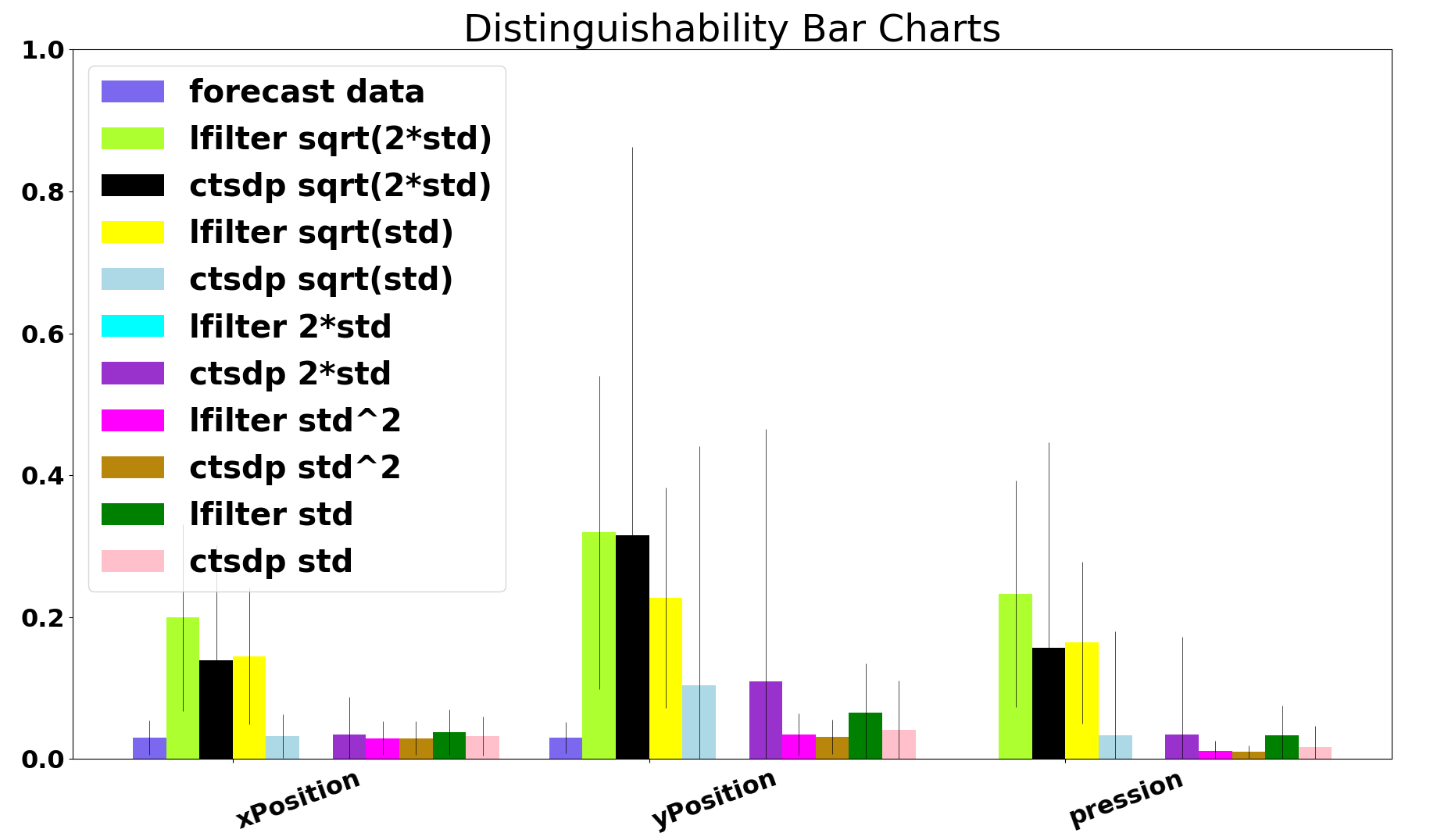}}
\end{figure*}
\begin{figure*}[!t]
\captionsetup[subfloat]{farskip=2pt,captionskip=1pt, font=scriptsize}
\tabcolsep=0.11cm
\centering
\subfloat[]{\label{feat_IG_HW}
      \includegraphics[scale=0.09, keepaspectratio]{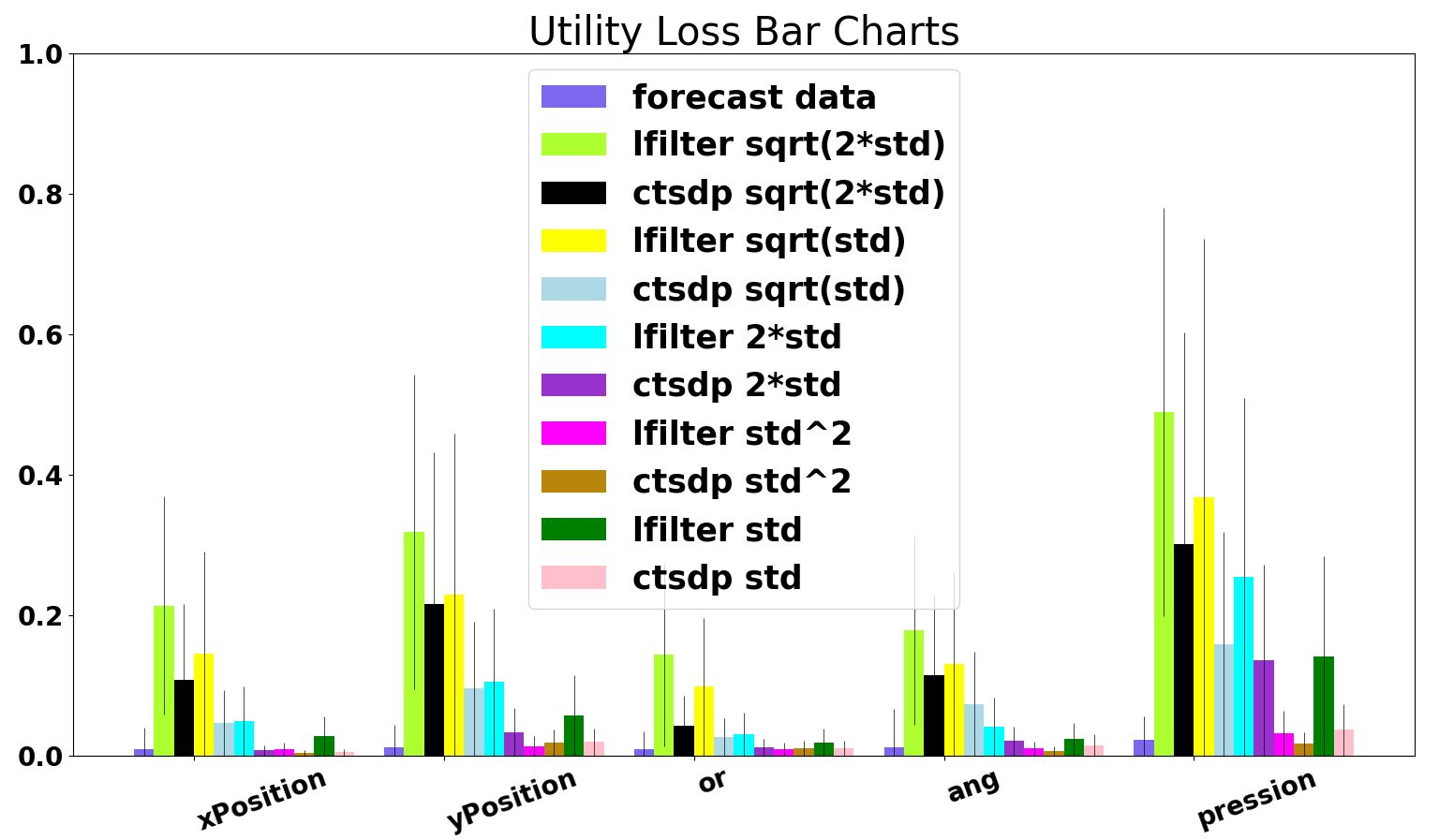}}
\subfloat[]{\label{feat_IG_Dig}
      \includegraphics[scale=0.09, keepaspectratio]{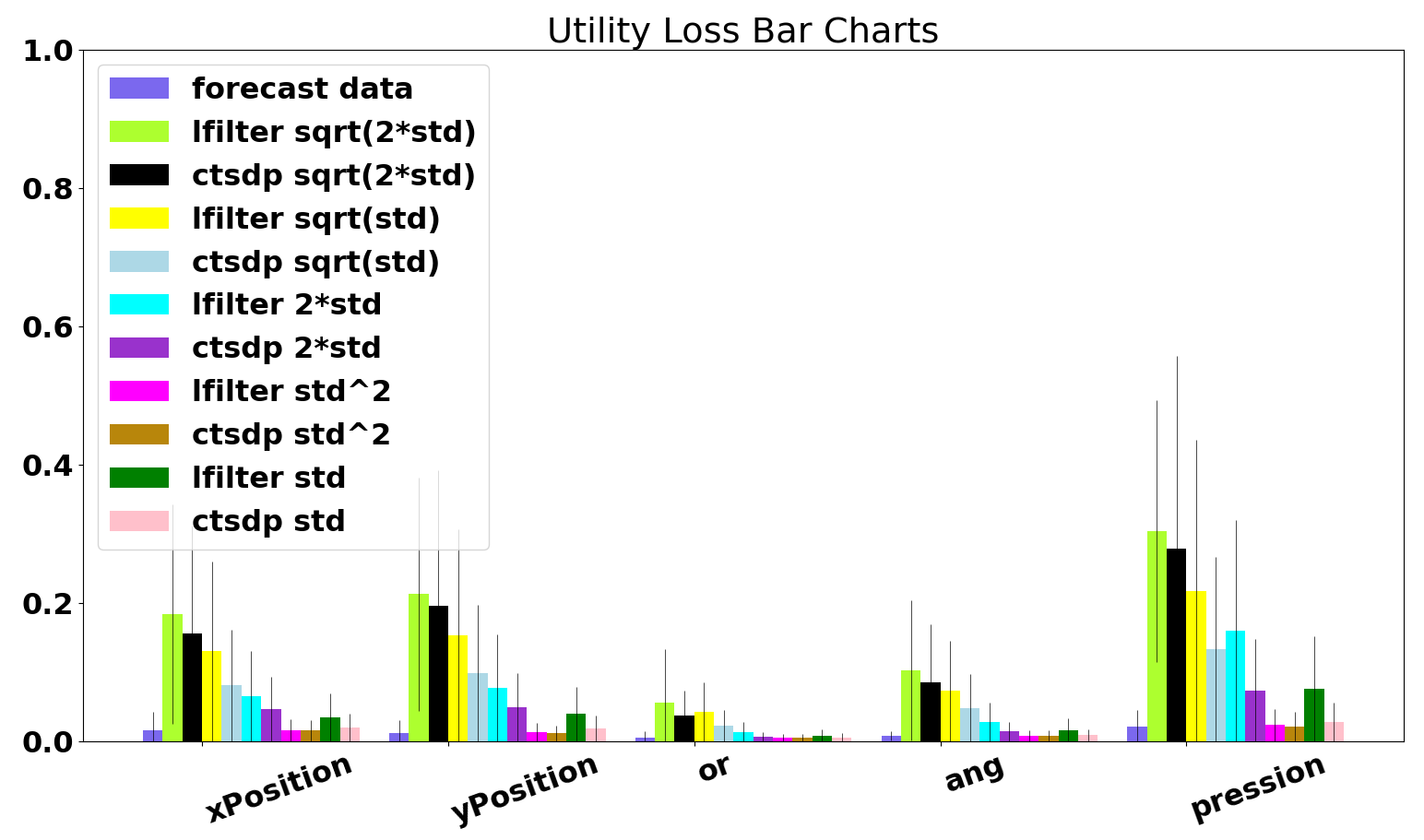}}  
\subfloat[]{\label{feat_IG_Swipes}
      \includegraphics[scale=0.1, keepaspectratio]{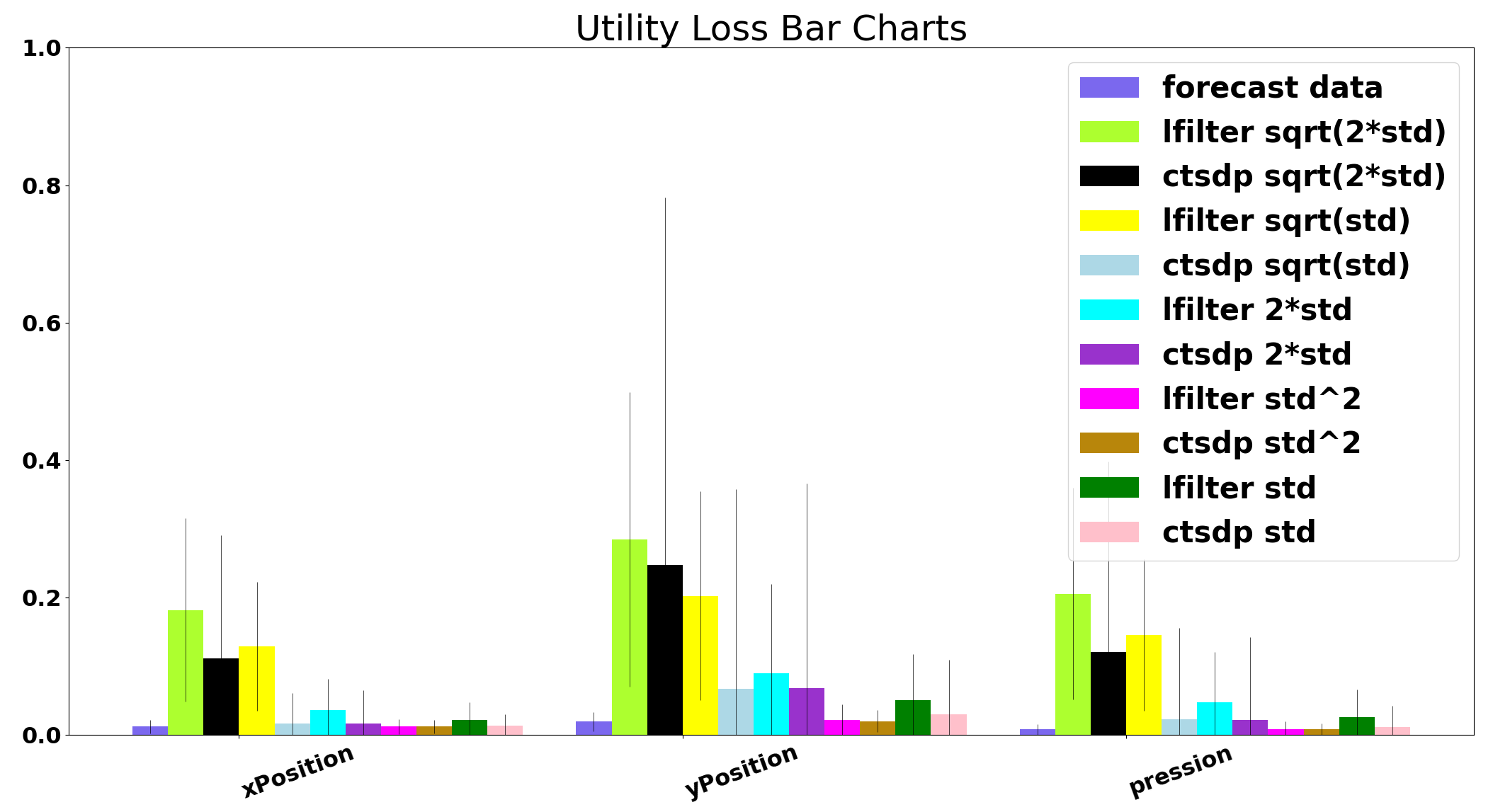}}
\caption{Effect of obfuscation mechanism on individual features. First row shows the results of Untrackability from all the datasets, second row shows the results of distinguishability from all the datasets, and third row shows the results of Utility Loss from all the dataset.}
\end{figure*}

Based on the above results, we select the best values of Gaussian noise scale and correlation threshold to discuss the effect of obfuscation on features and specific functionalities (i.e. letters, digits, and swipe). We perform experiments to various noise scale: square root of 2 times the standard deviation of the incoming series (sqrt(2*std)), square root of the standard deviation (sqrt(std)), 2 times the standard deviation (2*std), squared standard deviation (std\textasciicircum 2) and the standard deviation (std).

\subsection{Effect on Different Features}
\label{feat_res}
This subsection discusses the effect of our obfuscation mechanism on the individual features of the datasets. For the  HW(L) datasets, the highest untrackability and distinguishability are provided by \textit{Pressure} followed by \textit{yPosition} with all noise scales. However, with noise scale of sqrt(2*std), \textit{xPosition} offers high untrackability (Fig.~\ref{feat_untrack_HW}). These results indicate that \textit{Pressure} and \textit{yPosition} are effective against noise filtering attacks since high correlation means adversary is unable to filter noise from the original data. On the contrary, \textit{Pression} works best against untrackability because of high level of noise scale. We also notice that utility loss is highest with noise scale sqrt(2*std), which means that noise scale has more adverse impact on the utility than correlation threshold. 

For the HW(D) dataset, the highest untrackability is offered by \textit{pressure} of 49\% followed by \textit{yPosition} (30\%) with noise scale sqrt(2*std). However, with noise scale sqrt(std), \textit{xPosition} offers untrackability of 18\% (Fig.~\ref{feat_untrack_Dig}). We observe that \textit{pen angle} provides low distinguishability of 15\% with noise scale sqrt(2*std) and \textit{xPosition} offers distinguishability of 20\% with the same noise scale (Fig.~\ref{feat_IG_Dig}), respectively. 

Results from the Swipes dataset indicate that all features offer nearly equal amounts of untrackability rate with noise scale of sqrt(2*std). For instance, \textit{yPosition}, offers 37\% of average untrackability followed by \textit{Pressure} of 31\%. For distinguishability, both filtering methods (linear filter and CTSDP) give equivalent results for feature \textit{yPosition} when the noise scale is sqrt(2*std), in which case the distinguishability of 35\% using both methods (Fig.~\ref{feat_IG_Swipes}). 

\textbf{Remark 3:} We also analyze the effect of obfuscation on different gesture types, for example, letters, digits, or swipes in Setting A only, as this setting gives good results in terms of utility. Due to space limitation, we present these results in Appendix A of the supplementary material. We urge interested readers to go through supplementary material for detailed analysis of these results. 

\begin{figure*}[!t]
\tabcolsep=0.11cm
\captionsetup[subfloat]{farskip=2pt,captionskip=1pt, font=scriptsize}
\centering
\subfloat[HW(L)]{\label{fig:time_HW}
      \includegraphics[scale=0.09, keepaspectratio]{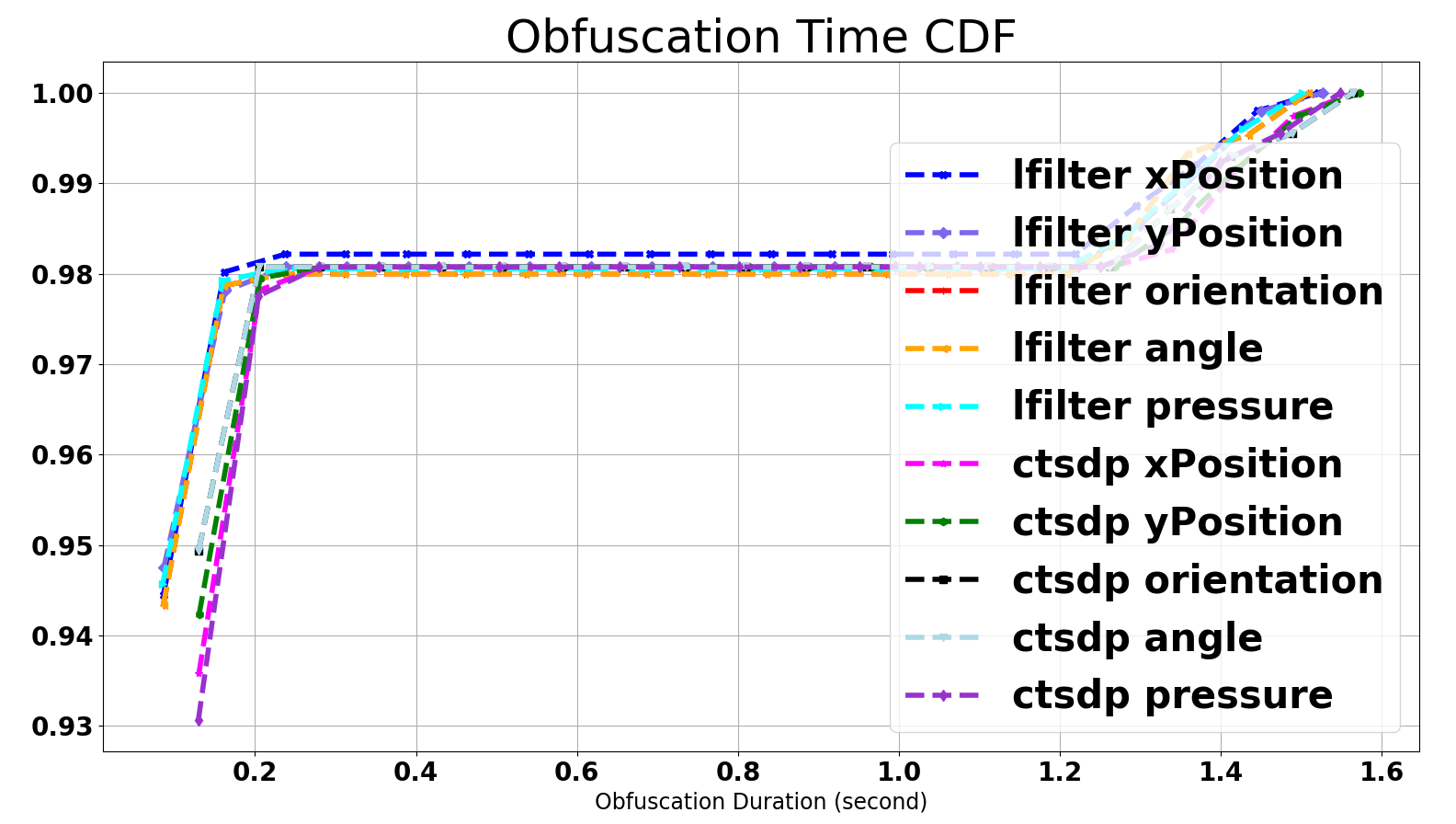}}
\subfloat[HW(D)] {\label{fig:Dig_time_HW}
      \includegraphics[scale=0.09, keepaspectratio]{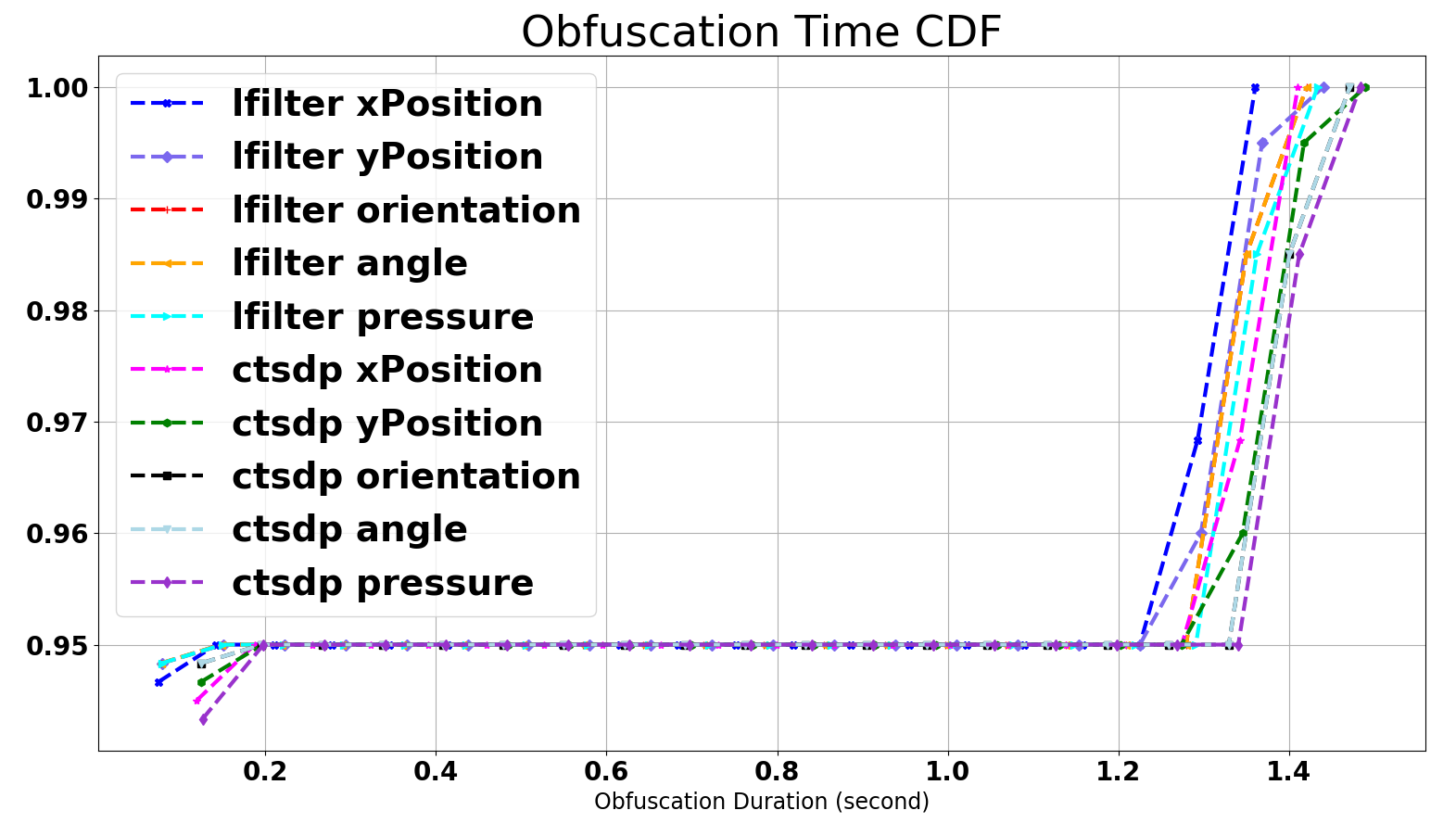}}
\subfloat[Swipes] {\label{fig:Swipes_time_HW}
      \includegraphics[scale=0.124, keepaspectratio]{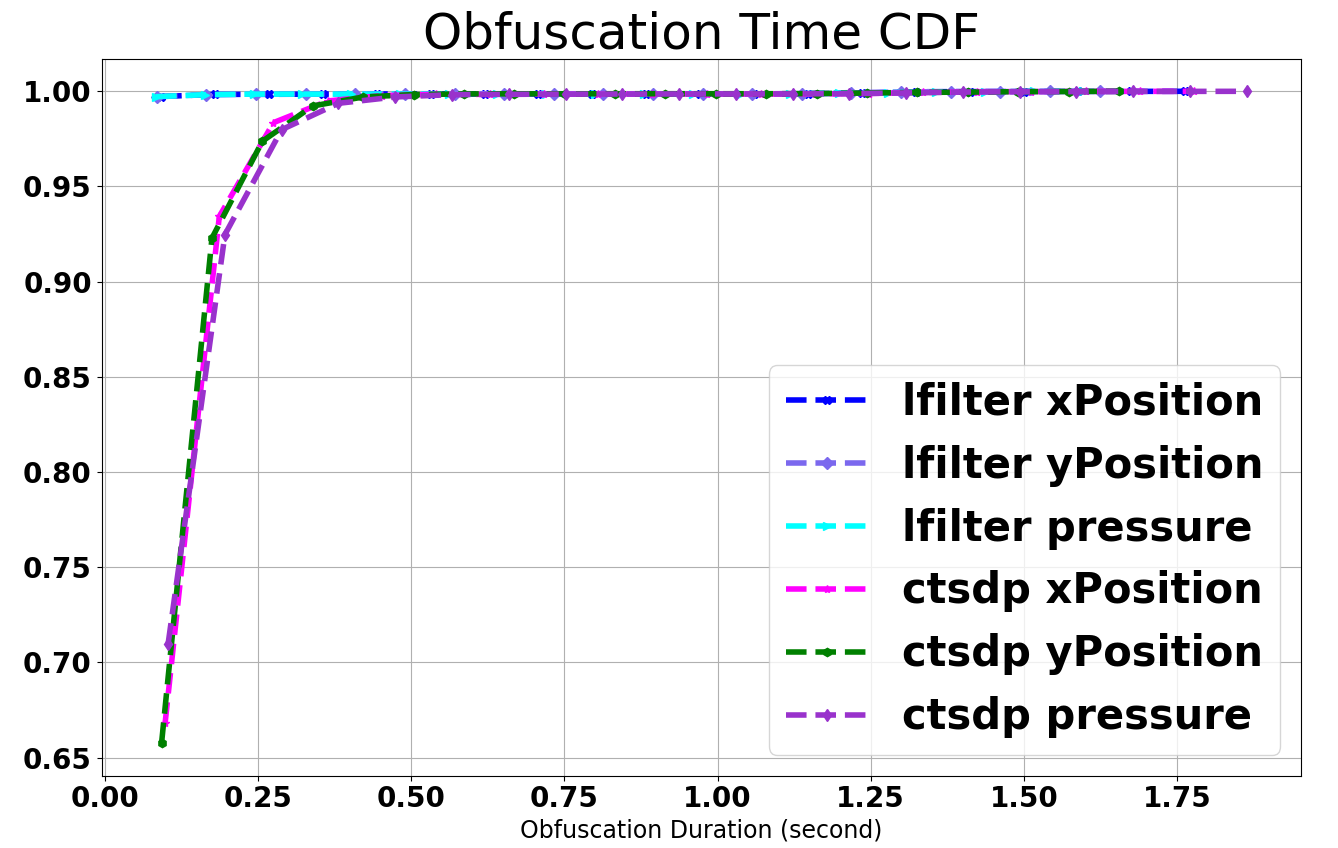}}
\caption{Obfuscation time required for different Datasets}
\label{feat-time}
\end{figure*}

\subsection{Execution Time}
\label{time_res}
We notice that our obfuscation mechanism comes with the cost of the overhead of training the mechanism and applying the obfuscation at run-time. The overhead of pre-trained models is reasonable as these models are pre-installed during the deployment phase of privacy mechanism. Furthermore, the overhead of updating these trained models can be avoided if updation is performed during off-peak hours i.e. when a user is not using the mobile device.

Fig. \ref{fig:time_HW} shows the time in seconds to obfuscate time-series data of HW(L) dataset at run-time. We find that 98\% of all time-series is obfuscated within 0.2 seconds, whereas it could take 1.6 second for the remaining 2\%. Figure \ref{fig:Dig_time_HW} shows obfuscation time of HW(D) dataset. Here, we find that 95\% of time-series are obfuscated in less than 0.2 seconds whereas the remaining 5\% takes an average of 1.4 second for obfuscation.
We observe that obfuscation time for swipes dataset is much less than the HW(L) and HW(D) datasets. The average obfuscation time to obfuscate all the features is less than 0.5 seconds for all time-series.

\textbf{Remark 4:} We also analyze the effect of different phases of the obfuscation mechanism on privacy (untrackability) and utility. The results are presented in Appendix A of the supplementary material. 

\section{Discussion}
\label{dis}

In this section, we discuss some key findings from our results and limitations of this study.

\begin{enumerate}

\item \textbf{Trade-off between Privacy against Noise Filtering Attack and other Privacy Risks:} Our obfuscation mechanism is capable of reducing the risk of noise filtering attack by an adversary as it adds noise which is correlated with the original data. However, prevention against adversary noise filtering attack needs to be tradeoff with other privacy aspects i.e. untrackability and indistinguishability. Therefore, the obfuscation mechanism employs two parameters ``Guassian Noise Scale’’ and ``Correlation Threshold’’ to balance between these two properties. Our results indicate that a good balance can be achieved where noise filtering attack can be reduced whilst other privacy properties are also guaranteed. 

\item \textbf{Comparison of Untrackability and Indistinguishability: }Our obfuscation mechanism is more effective against untrackability than the indistinguishability metric. For instance, a noise scale of 0.8, gives untrackability of 68\% for the GPS dataset whereas indistinguishability is limited to 22\% only. We find similar trend across other datasets as well. The low indistinguishability results are something which need further investigation and may require an extended version of the obfuscation algorithm. 

\item \textbf{Large Number of Samples Give More Effective Results:} As evident from the results of the GPS dataset, our obfuscation mechanism offers better privacy and utility guarantees with large number of samples. Thus, our obfuscation mechanism will be more effective when deployed at a large-scale. 

\item \textbf{Some Sensor Features are more Effective than Others:} Our results from Sec.~\ref{feat_res} show that some features are more effective in providing privacy guarantees than the other features and these features happen to have common results across all datasets. 

\item \textbf{Obfuscation Time Needs Improvement:} Since obfuscation is performed on-the-fly, the obfuscation time should be significantly short for such online processing. Our mechanism needs improvement in obfuscating data for efficient online processing. It is possible by optimizing implementation and utilizing efficient versions of forecasting and clustering algorithms. We aim to improve run-time efficiency of the proposed mechanism in future. 

\item \textbf{Obfuscation is Effective against in a range of Mobile Apps and Services:} Our results from different datasets indicate that the proposed obfuscation mechanism is effective across range of mobile apps and services that utilize sensor APIs. However, in future, we plan to evaluate our obfuscation method for other sensor data types such as motion sensors (e.g. accelerometer and gyroscope). 

\end{enumerate}
 
\section{Related Work}
\label{related_work}

A number of techniques have been developed that formulate obfuscation problem as a min-max optimization and solve using adversarial networks.  Raval et al. ~\cite{raval2019olympus} proposed utility-aware obfuscation framework, OLYMPUS, that limits the risk of disclosing user sensitive information from sensors data such as images. The framework was based on generative adversarial networks (GANS), with a game between obfuscator and an attacker. However, the framework is user-dependent where an input from a user is required to specify privacy and utility labels. Their threat model trusts a user/developer to hook up an app with a framework.  Moreover, the framework is applicable only to apps that are using machine learning classifier to output utility classification score.

Malekzadeh et al. ~\cite{malekzadeh2018replacement, malekzadeh2018protecting, malekzadeh2018mobile} proposed approaches to protect sensory data using autoencoders. However, their schemes only work when public (non-sensitive) and private (sensitive) data are clearly mentioned. Shokri ~\cite {shokri2015privacy} proposed a methodology for designing optimal user-centric obfuscation mechanisms against adaptive inference attacks. This approach works well on certain data types such as location trajectories that are discretized, however, it is not clear how this scheme works for continuous data such as images, touch sensors, GPS locations etc.

Das et al. ~\cite{das2016tracking} studied the feasibility of conducting sensor fingerprinting on mobile phones and also discussed countermeasures based on calibration and noise addition. In another paper, Das et. al ~\cite{das2018every} proposed an obfuscation scheme to defeat motion sensors based fingerprinting. Erdogdu ~\cite{erdogdu2015privacy} proposed a privacy-utility aware framework for time series data using information theoretic approach. However, these schemes do not clearly describe how their framework behaves with different sensor data types.

Our proposed obfuscation mechanism is significantly different from the above literature from many aspects. First, our mechanism is proposed for obfuscating the real-time data; it means that any sensitive data is obfuscated at the run-time before sending to the remote servers. Most of the existing literature on protecting the data using adversarial methods or deep neural network require offline batch processing. Second, our mechanism holistically covers different types of mobile sensor data such as touch sensors, motion sensors, GPS, etc. The existing mechanisms focused mostly on specific data type and did not discuss the suitability for the other types of data. Third, our method does not depend on users, developers, or a third-party to train or run the mechanism. It requires no user-interaction for the data labeling or data training.

\section{Conclusion}
\label{conclusion_mob_pri}

We have proposed a privacy preserving mechanism that protects users from the threats of information leakage (tracking and distinguishability) from the mobile sensor data without significant utility loss. The proposed mechanism overcomes the drawback of existing solutions in addressing fluctuations in sensor data, by utilizing the concept of time-series modeling and forecasting. Moreover, the mechanism works in isolation from user or service providers/app developers and is data-type independent. We use correlated time-series noise addition to overcome the threat of noise filtering attack by an adversary. We empirically evaluate the mechanism on three different datasets and show that the risk of trackability and distinguishability are reduced whilst preserving the functionalities of the app to a sufficient level.

\bibliographystyle{abbrv}
\bibliography{pets}  

\end{document}